\documentclass{PoS}
\usepackage{url}

\title{Summary of the 36$^\mathrm{th}$ ICRC Gamma-ray Indirect sessions}

\ShortTitle{}

\author{\speaker{Julian Sitarek}\\ 
        Department of Astrophysics, The University of Lodz, ul. Pomorska 149/153, 90-236 Lodz, Poland\\
        E-mail: \email{jsitarek@uni.lodz.pl}}

\abstract{
At the 36$^\mathrm{th}$ ICRC during 11 parallel Gamma-Ray Indirect sessions in total about 70 talks were presented. A few of the plenary highlight talks as well concerned mostly indirect observations of gamma rays. 
In addition about 140 posters on this topic have been shown in poster sessions. 
This rapport tries to summarize the results presented in those contributions.}

\FullConference{36th International Cosmic Ray Conference -ICRC2019-\\
		July 24th - August 1st, 2019\\
		Madison, WI, U.S.A.}

\begin{document}

\section{Introduction}
Gamma rays are the highest energy part of the electromagnetic spectrum.
As such they are a unique tool to probe the most energetic and the most extreme processes in the Universe. 
Classically the difference between high-energy (HE, $\lesssim 100$\,GeV) and very-high-energy (VHE, $\gtrsim 100$\,GeV) gamma rays is also reflected in the most optimal method for measuring them.
For HE gamma rays the most effective way of measuring them is to flight a detector on a satellite and directly measure the primary gamma rays falling into the detector.
On the other hand for VHE gamma rays the typical fluxes of such sources are low enough that they would require unrealistically large detector sizes.
In the latter case, the Earth's atmosphere becomes a part of a detector, which allows indirect measurement methods of gamma rays. 
A gamma ray hitting the atmosphere starts an electromagnetic cascade of secondary $e^-$, $e^+$ and gamma-ray photons, the so-called air shower.
Two separate indirect methods measure such cascades.
Surface arrays of detectors measure the secondary particles that reach the ground level.
Such an instrument has a high (close to 100\%) duty cycle and also a large field of view (FoV).
However their performance parameters, such as energy threshold and resolution as well as short term sensitivity are rather poor.
In contrary Imaging Atmospheric Cherenkov Telescopes (IACTs) measure the Cherenkov light produced by secondary charged particles in the air shower. 
They provide a complementary method for studying VHE gamma rays to surface arrays.
The observations of IACTs are mostly limited to good weather and dark nights resulting in a low duty cycle of $\sim 11\%$\footnote{The duty cycle can be somewhat increased by performing observations with a reduced performance under partial cloud coverage or during moonlight.}.
In addition the FoV of IACTs is usually only a few degrees across. 
However, their performance parameters, such as energy and angular resolution as well as short-term sensitivity and energy threshold allow in-depth studies of individual sources.

In these proceedings I report on the results presented in the conference that were obtained with those two techniques and on the interpretations of such results.
In Section~\ref{sec:inst} the instruments that mostly contributed to the results in the conference are briefly introduced. 
In Section~\ref{sec:gal} the progress in galactic physics is discussed. 
Similarly, in Section~\ref{sec:egal} the progress in extragalactic physics (mainly active galactic nuclei) is discussed. 
In Section~\ref{sec:trans} the multiwavelength and multimessenger observations of short transients are summarized. 
The search for emission from more exotic possible emitters of VHE gamma rays is presented in Section~\ref{sec:exotic}.
A few selected observational methods are discussed in Section~\ref{sec:methods}.
Final summary and outlook are given in Section~\ref{sec:sum}.

\section{Instruments}\label{sec:inst}
At the conference, results from various instruments were presented.
H.E.S.S. \cite{c656} is a system of 4+1 IACTs and the only one currently operating in the Southern hemisphere. 
In the Northern hemisphere two major IACT arrays are currently operating: MAGIC \cite{2016APh....72...61A} and VERITAS \cite{c632,c773}. 
In addition, a smaller telescope located on the MAGIC site, FACT \cite{c665}, is used for monitoring of bright blazars.
The next generation IACT experiment, the Cherenkov Telescope Array (CTA) Observatory \cite{c741} is currently in the prototyping and harmonization phase and will be constructed in two sites to cover both hemispheres. The first telescope of CTA, LST1 is already installed in the Northern CTA location \cite{c653}.  
Currently only the Northern part of the sky is covered by surface arrays (HAWC \cite{c015,c736}, and Tibet AS+MD array \cite{c778}).
A similar installation is also considered for the Southern hemisphere (SWGO, \cite{c785,c786}).
Some instruments combine both measurement techniques of gamma rays in a hybrid system. 
The construction of LHAASO \cite{c693} is expected to be finished by the end of 2020 and the first of the WCDA detectors was already able to detect 5 known gamma ray emitters. 

The above-mentioned, currently operating experiments have also been undergoing upgrades in the last few years to improve their performance and allow new observation modes. 
In particular the cameras of the four smaller H.E.S.S. telescopes were upgraded, which allowed a higher flexibility in the analysis \cite{c834}, lower dead time increasing the effective performance, and larger number of stereo events with the larger, central telescope. 
In the case of MAGIC a dedicated trigger (the so-called Sum-Trigger-II, \cite{c802}) has been developed for the lowest energies, while a special observation mode at high zenith angles has been used for the highest energies \cite{c828}. 
The VERITAS Collaboration has invested into developing special hardware for allowing optical intensity interferometry \cite{c714}. 
The HAWC Collaboration constructed an outrigger array of tanks surrounding the main array \cite{c736} to improve the array performance, in particular at higher energies. 

\section{Galactic physics}\label{sec:gal}
\begin{figure}[t]
  \centering
  \includegraphics[width = 0.9\textwidth]{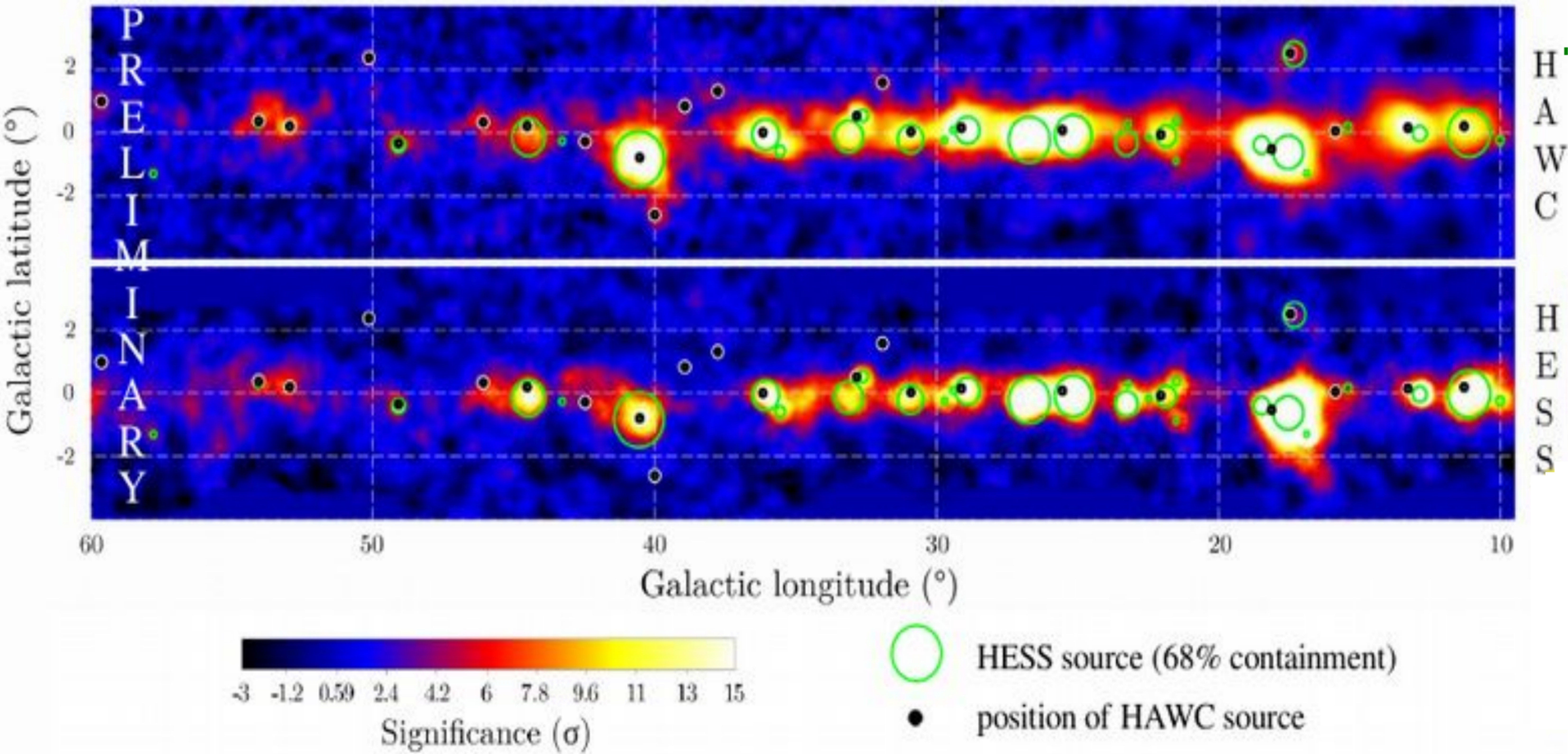}
  \caption{Top panel: HAWC Galactic plane survey. Bottom panel: H.E.S.S. Galactic plane survey using $0.4^\circ$ correlation radius and field of view background method, adopted from \cite{c706} presentation.}\label{fig:hess_hawc_gps}
  \end{figure}

Despite the small FoV limiting the survey capabilities, one of the most important contributions of IACTs to Galactic TeV gamma-ray astronomy is the Galactic plane survey performed by H.E.S.S. with almost 2700\,h of data \cite{c697}. 
Now those results can be compared with the Galactic plane observations performed with an alternative technique, by HAWC.
The original differences in the observed morphology of sources and most of the differences in the detections of individual sources are instrument-driven, in particular due to the effect of the background estimation \cite{c706} (see Fig.~\ref{fig:hess_hawc_gps}).
The Galactic Plane Scan allowed probing the population of various types of TeV sources in our Galaxy.
Currently about 150 such sources are known \cite{tevcat}.
The modelling of the source luminosity function suggests that there is still an order of magnitude more sources remaining to be detected and constituting 50\% of the flux of already known sources \cite{c801}. 

The observations of the Galactic center by H.E.S.S. resulted in a detection of a diffuse emission with a $1/r$ profile, which suggests a diffusion of Cosmic Rays (CR) injected in a quasi-continuous way from the central source \cite{2016Natur.531..476H}. 
This result has been recently confirmed by MAGIC \cite{c680} using an alternative morphology analysis method. 
The interpretation of those results is however not that straightforward. 
Instead of a single source accelerating protons up to PeV energies, the emission might arise from the inhomogeneous Galactic Cosmic Ray sea \cite{c816}. 

\subsection{Pulsars, their nebulae and halos}
Pulsars are rapidly rotating neutron stars.
The wind of $e^+e^-$ particles emitted from the pulsar is causing the formation of the Pulsar Wind Nebula (PWN). 
During the conference a detection (with $4.7\sigma$ significance of the excess) of a new pulsar PSR B1706-44 \cite{c799} by H.E.S.S. has been announced, and the MAGIC detection of Geminga \cite{c728}  pulsar has been shown, increasing the catalog of the IACT-detected pulsars to 4 sources.
Further searches have been performed in 13 archival pulsars observed by VERITAS, but no additional source has been detected \cite{c773}. 
Already within this small population some differences can be seen.
While the Crab pulsar has a spectrum extending up to TeV energies, both Geminga and PSR B1706-44 were detected only at the edge of the \textit{Fermi}-LAT spectrum (at the energies of tens of GeV).
In the case of the Vela pulsar, the detection is also at tens of GeV however,  multi-TeV excess has been previously claimed (not presented at the $36^\mathrm{th}$ ICRC conference)

\begin{figure}[t]
  \centering
  \includegraphics[width = 0.49\textwidth]{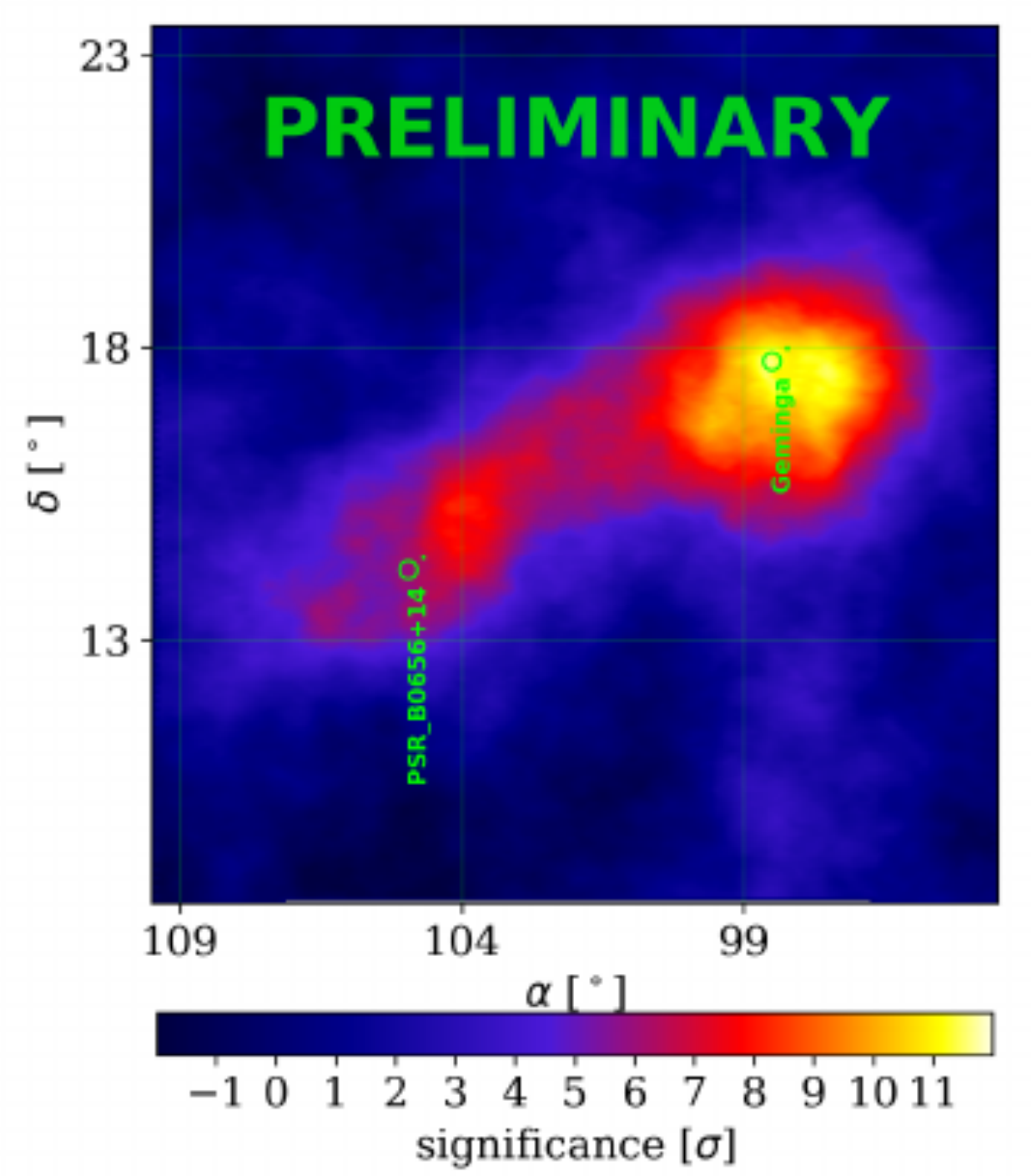}
  \caption{HAWC significance maps of the region of:
    Geminga and Monogem (green markers) smoothed by a $2^\circ$ top-hat function (left panel, \cite{c832}),
    2HWC J1825-134 region (right panel, \cite{c781}) . 
  }\label{fig:halos}
  \end{figure}

One of the recent surprising results was the detection of the so-called halos around the two neighbouring pulsars: Geminga and Monogem  \cite{2017Sci...358..911A}.
They suggest that the $e^+e^-$ leak out from the pulsar neighbourhood into a region of a 100 times smaller diffusion coefficient than the average one of the Galaxy.
The increased dataset presented by HAWC \cite{c832} (see Fig.~\ref{fig:halos}) allowed probing the energy dependent morphology.

There is however no commonly accepted explanation of the halos, with a few competing models possible. 
The halo emission might come from cosmic ray self-containment or from strong interstellar turbulence in the disk \cite{c020}. 
Another possibility is that Geminga PWN is still inside the old, not observable anymore supernova remnant (SNR) \cite{c670}. 
In addition, the correlation length at which the magnetic field varies might affect strongly the estimation of the diffusion coefficient \cite{c685}. 
In order to explain the phenomenon better it would be desirable to detect such halos around other pulsars. 
However, the current search for the emission from plausible candidates did not result in the detection of further sources \cite{c797}.
Using a similar morphology model of the emission to other pulsars (PKSB0540+23 and PSR J0633+0632) the diffusion coeffients have been studied, however the sample is too limited for a firm conclussion \cite{c640}. 

Currently 36 PWNe are known to emit in VHE gamma rays \cite{tevcat}. 
The observations provide both the spectral and morphological information that should be explained by the modelling. 
In particular, many of the H.E.S.S.-detected PWNe exhibit an offset between the pulsar position and the centroid of the PWN. 
Such offsets might come from the initial pulsar kick back in asymmetrical explosion, from propagation of the ejecta in inhomogeneous ambient medium, or from asymmetrical outflows of ions and leptons from the pulsar. 
Modelling of such morphologies require MHD simulations of pulsar winds, PWN and SNR \cite{c809}. 

H.E.S.S. observations of the HESS J1825-137 PWN revealed strongly energy-dependent morphology, such that rules out a pure diffusion model \cite{c715}.
Hints of energy dependent morphology have been also seen in Dragonfly PWN \cite{c639}.
The spectrum observed by HAWC is close to the one measured previously by VERITAS (some discrepancy most probably comes from a different integration region).
The same source has also been followed by the MAGIC telescopes exploiting the very high zenith angle observations technique and modelled as an extended source \cite{c827}.
Both instruments managed to measure the spectrum up to a few tens of TeV.

Observations of the 2HWC J1825 region have been reported with HAWC \cite{c781}.
The region is complex (see the right panel of Fig.~\ref{fig:halos}) and contains PWN, SNR and molecular clouds. 
The VHE gamma-ray emission could be disentangled into two sources and a hint of energy dependent morphology has been seen.
Even while the sources were modelled with leptonic (preferred) and hadronic scenarios the origin of the VHE gamma-ray emission is still not known \cite{c781}.

\subsection{Supernova remnants}
HAWC performed a search for SNRs using 3 years of data.
Out of the 9 sources selected on basis of their GeV detection, the HAWC visibility criterion and the lack of possible source confusion, 3 SNR were detected \cite{c674}.
All three are consistent with known TeV sources (W51, IC 433, $\gamma$ Cygni).

G106.3+2.7, a VERITAS-detected SNR, has been studies with the Tibet AS+MD array, resulting in detection of emission above 10 TeV \cite{c778}, however the spectrum is still under reconstruction.

HESS J1912+101, a bright unassociated TeV source with a shell morphology has been a target of a deep observational campaign with MAGIC and \textit{Fermi}-LAT.
Curiously, while the MAGIC observations, consistently with the earlier H.E.S.S. measurement, prefer a projected shell, at GeV energies the source can be better described with a radially symmetric Gausian \cite{c564}. 

\subsection{Search for PeVatrons}\label{sec:100TeV}
Cosmic rays of energies up to $10^{15}$ eV are thought of being of Galactic origin.
To understand the sources accelerating these hadrons up to this energy, gamma rays are an excellent tracer of their interactions.
Since protons would produce in hadronic interactions gamma rays with about one order of magnitude lower energy, searching for sources emitting gamma rays above 100 TeV can point us to PeVatrons.
However, gamma rays with such energies can also be produced in leptonic processes.
Thus, in order to prove that a source is indeed a PeVatron it is important to measure its spectrum with sufficient accuracy that would allow modelling in both leptonic and hadronic scenarios.
To obtain however good statistics of photons at such energies, where fluxes are very low,  a combination of large effective area and long observation time is required. 
In the conference, multiple contributions were devoted to searches for such sources, either with surface arrays that are natural candidates for long-term observations of the highest energies, or in the case of MAGIC with the very-high-zenith observations technique.

\begin{figure}
  \centering
  \includegraphics[width = 0.45\textwidth]{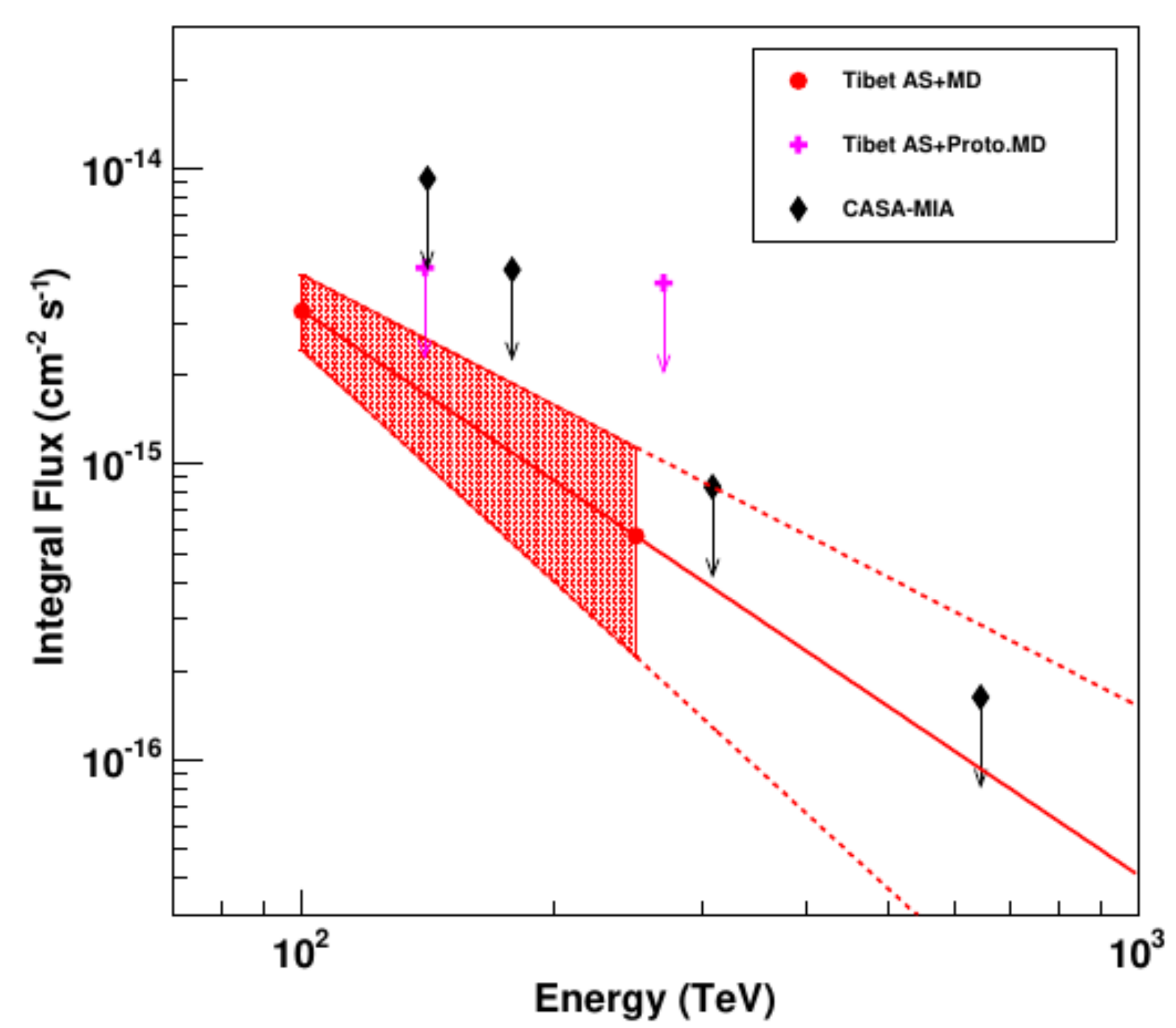}
  \includegraphics[width = 0.54\textwidth]{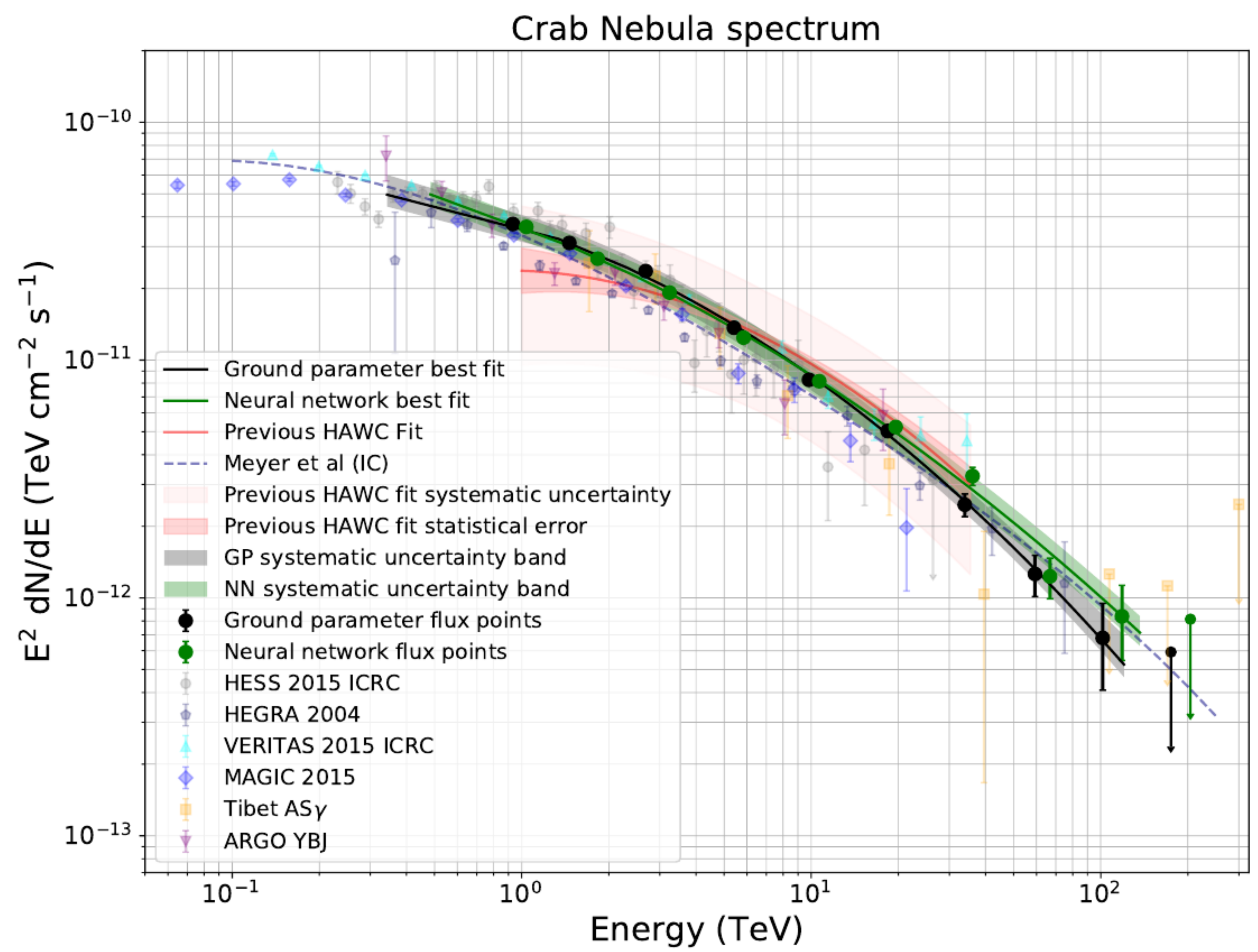}
  \caption{Crab spectrum observed with: Tibet AS+MD array (left panel, \cite{c712}) and HAWC (right panel, \cite{c723}).}
  \label{fig:crab_100tev}
\end{figure}

A natural candidate for looking for $> 100$\,TeV emission is the Crab Nebula, the strongest stable source in the VHE gamma-ray sky.
In addition, a long lasting disagreement of the highest energy part of the Crab spectrum measured by HEGRA and H.E.S.S. \cite{2004ApJ...614..897A,2006A&A...457..899A} makes it even more interesting to investigate emission at those energies.
All three instruments MAGIC \cite{c759}, HAWC \cite{c734} and Tibet AS+MD \cite{c712} array reach (or go beyond) energies of 100\,TeV without any visible cut-off (See Fig.~\ref{fig:crab_100tev}).
Individual photons with estimated energies of a few hundred TeV were reported, however the energy resolution as well as possible misclassified events should be taken into account when interpreting those.

HAWC performed a search of other sources emitting above 100\,TeV \cite{c734}. 
Three additional sources have been detected above 100\,TeV, however proving if they are indeed PeVatrons will require careful modelling of the emission.
It is interesting to note that those sources are close to radio pulsars, pointing to PWN or pulsar halo origin, and hence leptonic rather than hadronic origin of the emission  \cite{c734}.

\subsection{Binary systems and microquasars}

\begin{figure}
  \centering
  \includegraphics[width = 0.95\textwidth]{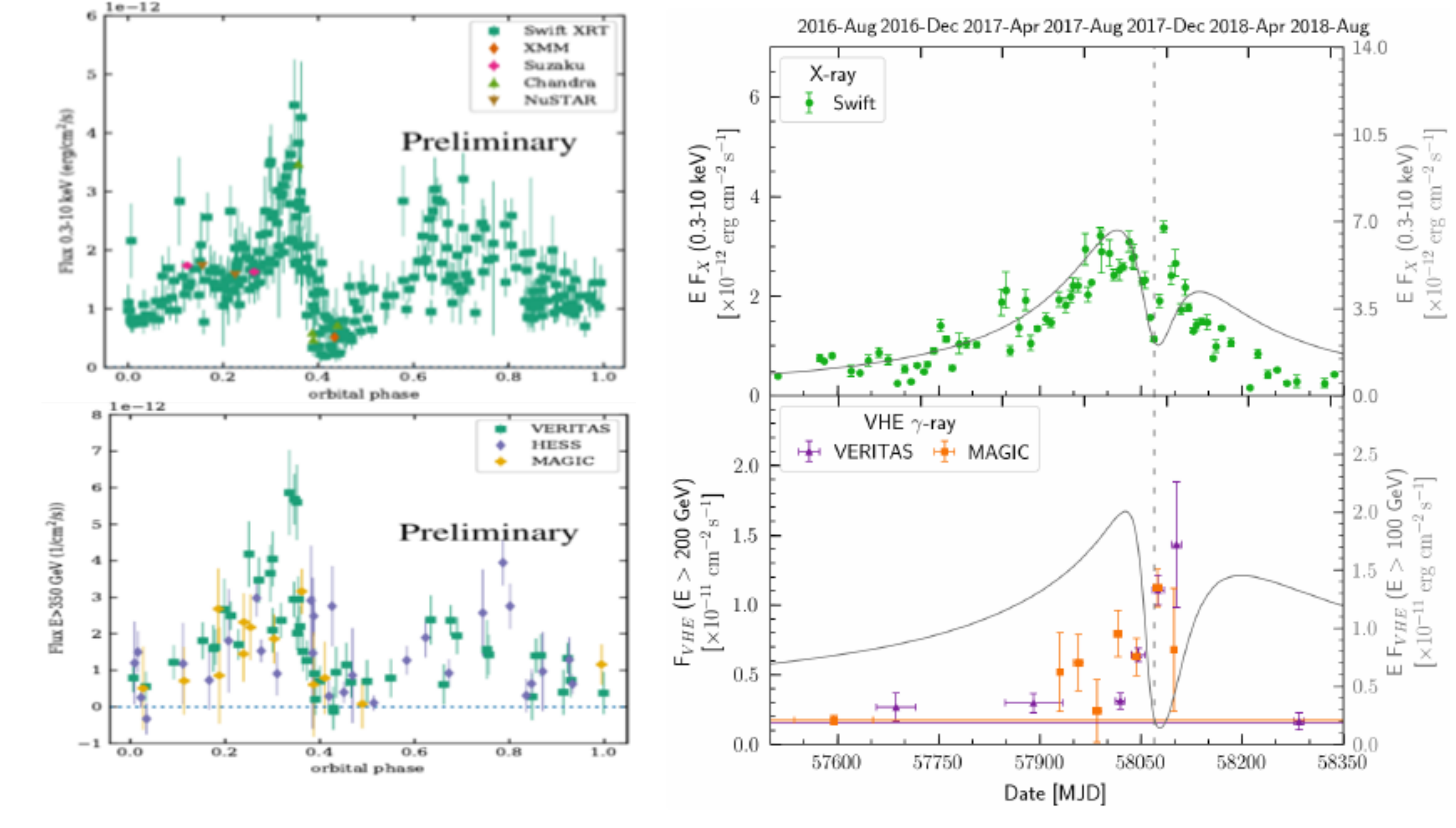}
  \caption{Phase folded light curve of HESS J0632+057 (left panels, \cite{c732}) and PSR J2032+4127/MT91213 (right panels, \cite{c824}).
  X-ray flux is shown in top panels, gamma ray flux in bottom. }
  \label{fig:binaries_lc}
\end{figure}

Binary systems are the only galactic sources in which mid- and long-term variability of TeV emission can be observed, often (but not always) correlated with the changes of the geometry as the two components of the system move along their orbits.
Therefore in order to understand such objects long term monitoring is essential.
In the case of binary systems with mid-scale periods such as HESS J0632+057 (a period of about a year) \cite{c732} or LS I +61 303 (period of about a month, \cite{c713}) an over dozen years of monitoring allowed H.E.S.S., MAGIC and VERITAS Collaborations to probe emission in multiple periods.
The rich data set allows a search for correlation of X-ray and gamma-ray emission and to search for features in the time dependence of the emission.
The gamma-ray as well as X-ray light curve of HESS J0632+057 shows a clear dip at phase $\sim 0.4$ (see left panel of Fig.~\ref{fig:binaries_lc}). 

PSR J2032+4127/MT91213 is a binary with a period of 50 years.
In November 2017 the periastron passage gave a possibility for ``once in a lifetime'' observation of emission in this phase \cite{c824}.
Long period allowed gathering of a large amount of MWL data to probe precisely the emission.
Contrary  to the case of HESS J0632+057, the X-ray and gamma ray emission of emission of PSR J2032+4127/MT91213 show completely different shapes, unable to be fully described by the current model.
In particular the rapid increase of X-ray flux without corresponding increase of VHE gamma-ray emission might come from the interaction between the pulsar and circumstellar disk of the companion star (see \cite{c824} and references therein).

\begin{figure}
  \centering
  \includegraphics[width = 0.95\textwidth]{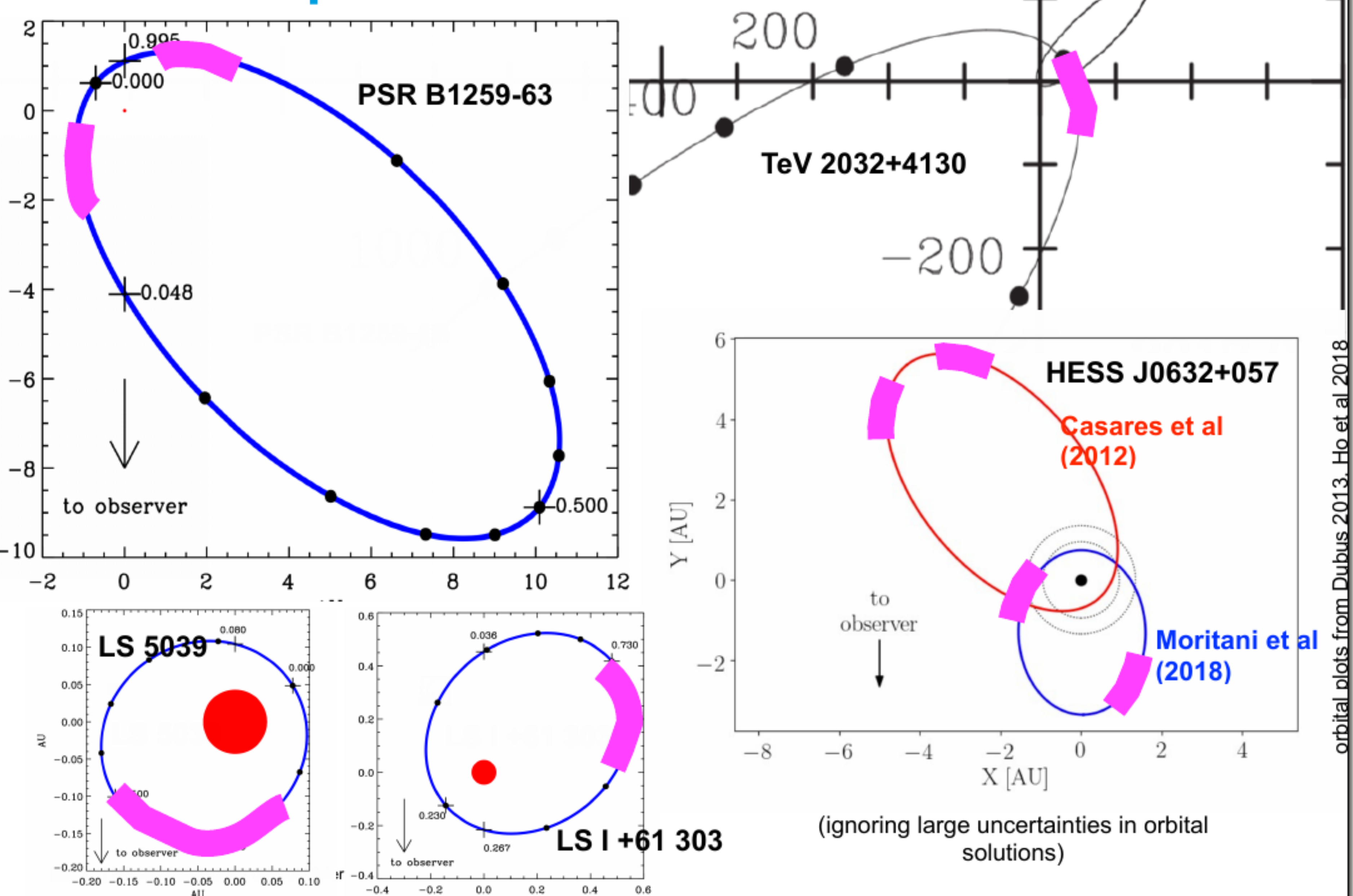}
  \caption{Position of the maximum (magenta regions) of the VHE gamma-ray emission in different binary systems (presentation of \cite{c732}). 
}
  \label{fig:binary_phase}
\end{figure}

It is remarkable to note that we still do not reach a common picture of gamma-ray emission from binary systems.
Even the phase at which the maximum of gamma-ray emission is observed varies between different binaries (see Fig.~\ref{fig:binary_phase}) with additional complication coming from uncertainty in the orbital parameters of binaries \cite{c732}.
The parameters of an orbit can be determined classically from the gas kinematics (see e.g. \cite{c813}), but also from the analysis of the gamma-ray light curve \cite{c1178}.
The latter method is however model dependent and therefore using such determined parameters in a modelling could result in biases. 

HAWC observations of microquasar SS 433 revealed TeV emission from the jet interaction regions \cite{c772}. 
SS 433 is the first microquasar in which the morphology of the TeV jets could be directly probed. 
Since the emission from both regions is seen as point-like, showing no evidence for a diffusion of long-lived protons, a leptonic scenario of the emission is favoured. 
\begin{figure}
  \centering
  \includegraphics[width = 0.4\textwidth]{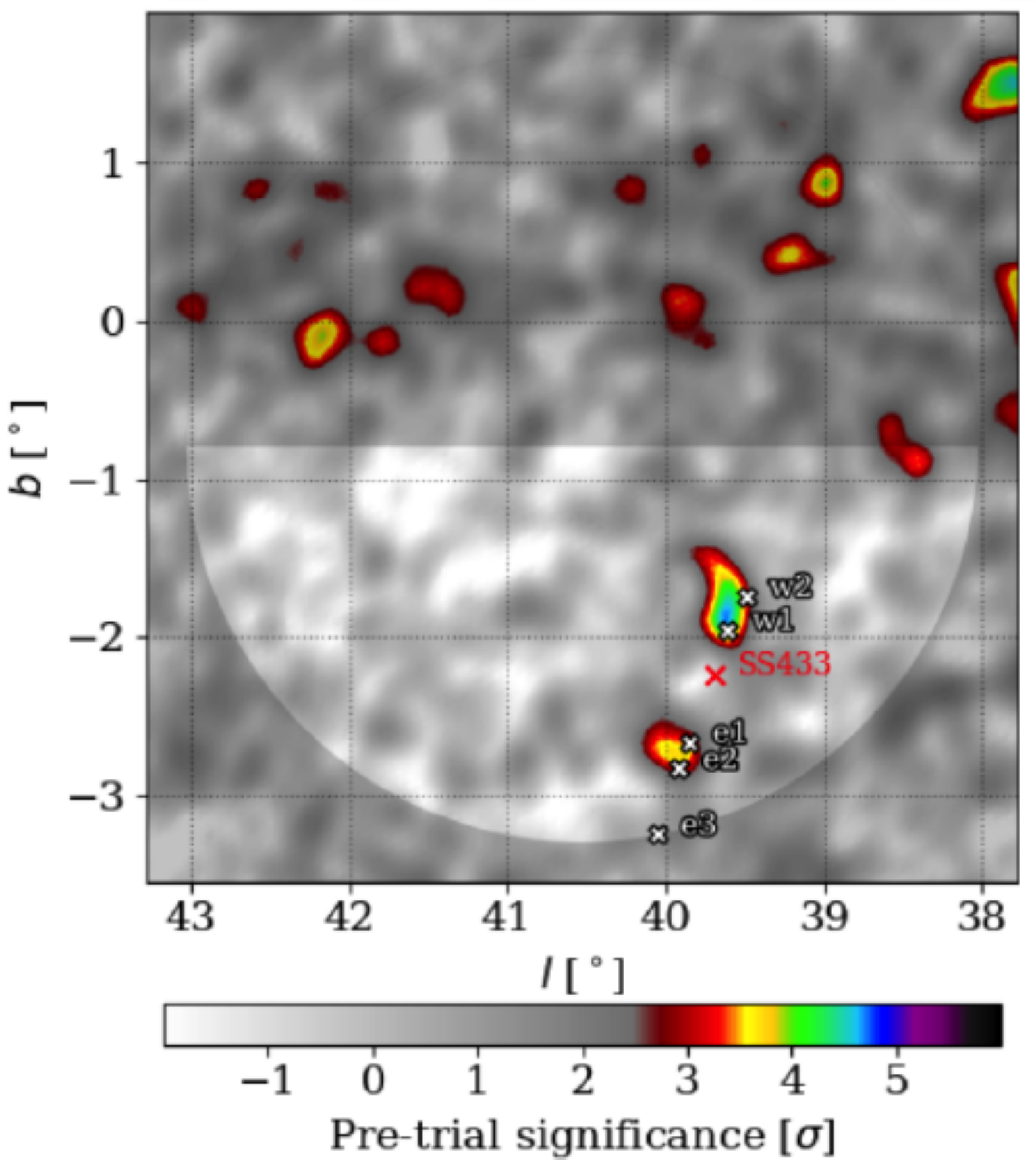}
  \includegraphics[width = 0.54\textwidth]{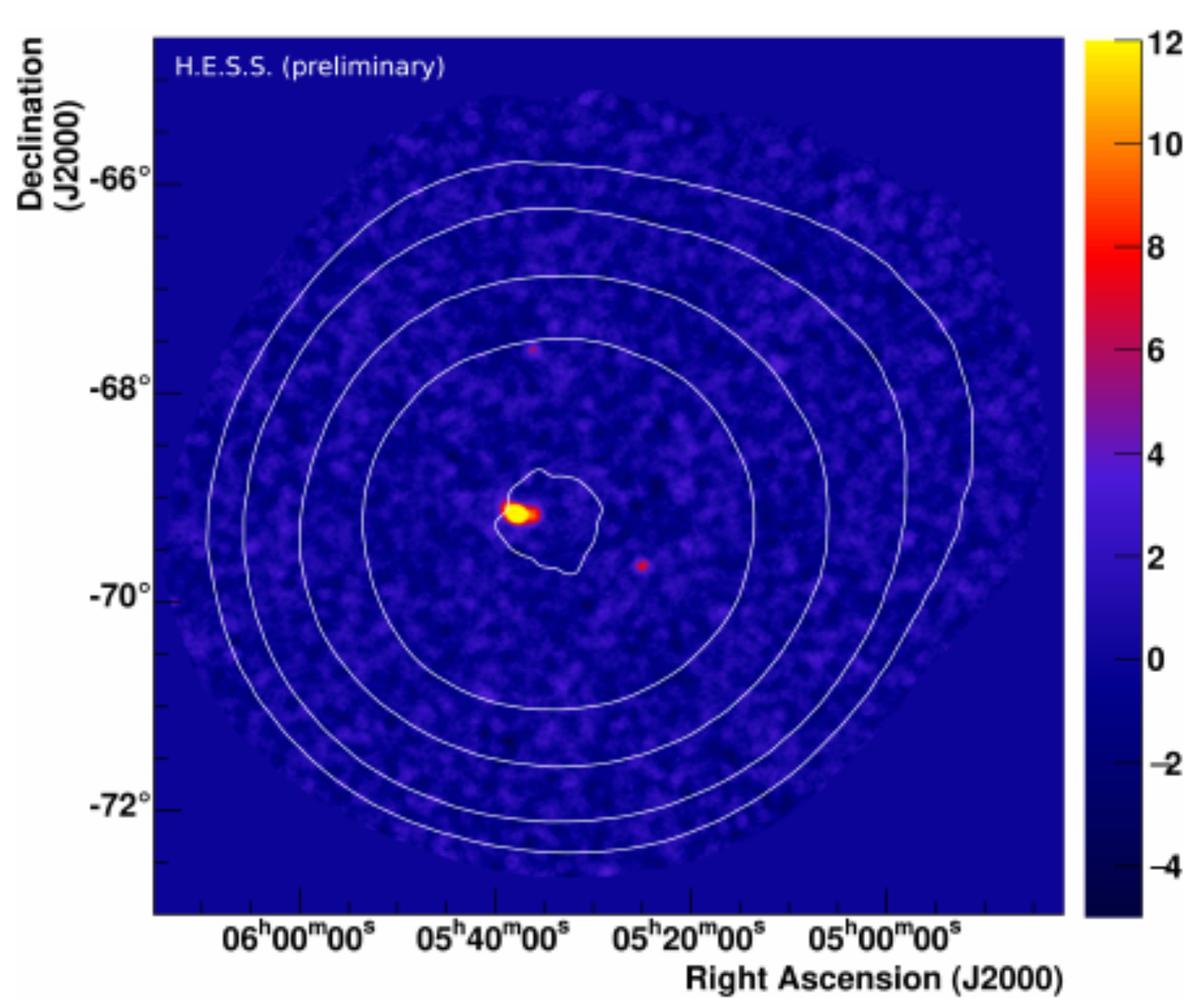}
  \caption{Left panel: The significance map of MGRO J1908+06 region by HAWC after subtracting the central source \cite{c772}. In the bottom part two interaction regions of SS 433 can be seen.
  Right panel: VHE gamma ray skymap of LMC \cite{c716}.}
  \label{fig:hawc_ss433}
\end{figure}

\subsection{Other Galactic sources}
One of the more complicated parts of our Galaxy is the Cygnus region which consists of a number of stellar clusters and associations.
Previous VERITAS observations showed also a complicated morphology of a VHE gamma-ray source in this region, consisting of a PWN and a Cocoon.
HAWC observations revealed that the cocoon emission extends beyond 100\,TeV and shows a spectral break with respect to the \textit{Fermi}-LAT spectrum \cite{c699}. 
The CRs in the cocoon could have originated in the OB2 association. 

H.E.S.S. observations of the Large Magellanic Cloud (LMC) allowed for the first time to study the physics of ``galactic'' sources outside of our Galaxy.
The current catalog of sources inside LMC consists of a PWN N 157B, superbubble 30 Dor C, SNR N 132D and a binary system LMC P3 (see \cite{c716} and references therein), showing rich family of types of sources, similarly to our Galaxy. 
Even while no new source in LMC had been claimed in this conference, the first VHE gamma-ray skymap of another galaxy (see Fig.~\ref{fig:hawc_ss433}, right panel) has been presented together with limits on a few tens of individual sources \cite{c716}.

\section{Extragalactic physics}\label{sec:egal}
Synergies between the IACT technique and observations with surface arrays show up also in the investigations of processes happening in  extragalactic objects. 
The catalog of extragalactic sources detected in VHE gamma-rays consists mainly of Active Galactic Nuclei (AGN), mostly blazars \cite{tevcat} in which the jet is pointed towards the observer. 
On one hand, the violent variability of those objects favours observations with instruments with high instantaneous sensitivity, i.e. IACTs, for in-depth studies of AGN flares. 
Also, the absorption of TeV gamma rays by the extragalactic background light (EBL\footnote{EBL is the integrated over time emission of all the stars and dust in the Universe}) for more distant sources again favours IACTs as they usually have lower energy threshold than surface arrays. 
On the other hand, the variable nature of AGNs requires constant monitoring, that is observational time consuming (and good-weather dependent) with IACTs, or follow-up of alerts from lower energies. 
In this case surface arrays can provide unbiased monitoring. 
Also surface detectors can provide flare alerts, however only for the brightest sources and flares. 

\subsection{Monitoring of AGN}
The HAWC Collaboration presented results from three years of a common monitoring with the \textit{Swift} satellite of a classical high energy peaked BL Lac (HBL) object Mrk 421 \cite{c682}. 
Such long-term monitoring allows studies of a X-ray to VHE gamma-ray correlation on a global, rather than flare-by-flare basis.
The FACT telescopes also perform long term monitoring of bright blazars with an impressive amount of $\sim 10 000$\,h observation time collected so far.
The monitoring covered 30 individual flares observed together with \textit{Swift} and showed a strong correlation of both fluxes without any significant time delay \cite{c796}. 

The long-term monitoring of blazars allows one also to search for possible signatures of periodic emission.
Curiously, the FACT monitoring of another bright blazar Mrk 501 has revealed a hint of quasi-periodic behaviour with a period of 332-days \cite{c665}.
Even while no statistical significance of the hint, or possible systematics effects have  been evaluated yet, it is interesting to note that a similar period has been obtained in the analysis of \textit{Fermi}-LAT data of the same source \cite{2019MNRAS.487.3990B}.

VHE gamma-ray monitoring of non-HBL sources is difficult due to their soft emission in this energy range.
The MAGIC monitoring of the Flat Spectrum Radio Quasar (FSRQ) PKS1510-089 performed between 2012 and 2017 showed that the source exhibits persistent VHE gamma-ray emission also during low GeV states \cite{c629}. 

\subsection{Extreme HBLs}
Extreme HBLs (EHBLs) lie at the highest energy end of the blazar sequence.
They are the weakest in flux, however their synchrotron peak is shifted to higher energies than for HBLs, peaking above $10^{17}$ Hz. 
Only a handful of such sources is known in VHE gamma rays. 
An effort to increase the VHE gamma-ray EHBL population was done by the MAGIC Collaboration by observing multiple promising targets. 
As a result 4 EHBLs were detected and a fifth one showed a hint of emission \cite{c768}. 
While the broadband emission of those objects can be explained in the framework of either spine-layer, synchrotron-self-Compton (SSC) or lepto-hadronic hadronic, the last two models require large deviation from energy equipartition (balance between the energy density of the magnetic field and relativistic particles) \cite{c768}.

\begin{figure}
  \centering
  \includegraphics[width = 0.9\textwidth]{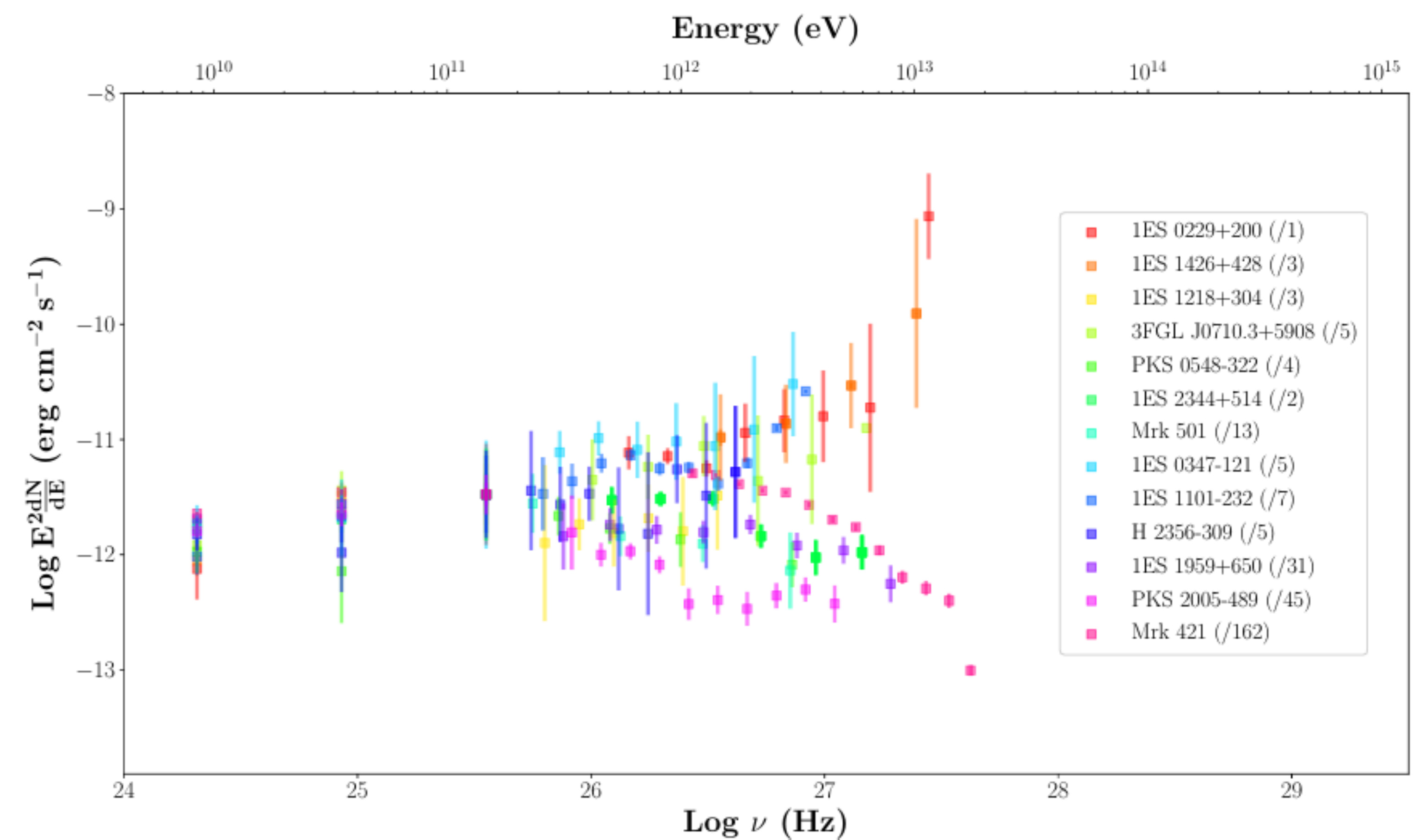}
  \caption{GeV-TeV spectral energy distribution of EHBLs, corrected for EBL absorption and normalized at 147GeV  to 1ES0229+200 flux \cite{c676}.}
  \label{fig:ehbl}
\end{figure}

With the population of known EHBL sources growing it becomes possible to answer the question whether they form a single population or if there are intrinsic sub-classes.
A natural question is if the extreme location of the synchrotron bump is also reflected in the shape and peak energy of the high energy bump in those sources. 
Fig.~\ref{fig:ehbl} shows normalized and corrected for EBL absorption spectra of 18 known in VHE gamma rays EHBLs. 
While the GeV spectrum is similar between the different sources, a large divergence is seen in the VHE band \cite{c676}. 

What complicates the EHBL picture even more is the typical blazar variability, which might even change the classification of a source. 
In the case of 1ES2344+514, the source was first detected in VHE gamma rays during a high flare in which the synchrotron peak position was above $3\times10^{18}$\,Hz, classifying the source as EHBL, however further observations of the source in low states showed no further EHBL behaviour (\cite{c620} and references therein). 
Nevertheless, a strong flare of 1ES2344+514 in 2016 showed a renewed extreme behaviour \cite{c620}.

Nearby EHBLs are also plausible targets for observations with surface arrays.
Observations of an archetypal EHBL 1ES0229+200 with HAWC, while did not result in a detection, provided flux limits that are in tension with the observations of this source with IACTs \cite{c822}.
This might be caused by the variability of the source.
Independently from the possible variability, the HAWC measurements strongly constrains the model in which the 1ES0229+200 VHE gamma-ray emission would be coming from interactions of the UHE CRs along the line of sight \cite{c822}. 
A dedicated search for variability in 6 EHBLs was performed with the VERITAS telescopes, however no variability has been observed in this sample \cite{c689}.

\subsection{Blazar flares}
Nearly all the extragalactic VHE gamma-ray sources belong to the class of blazars, mostly of the BL Lac type.
As blazars are extremely variable, they often give us a possibility to study in detail the emission processes in them during large outbursts when observed fluxes are high enough for precise measurements.
Nevertheless, randomness of this variability makes it difficult to catch such states with a broad coverage of instruments operating at different wavelengths. 

A hint of a curious narrow spectral feature has been observed by the MAGIC telescopes in the spectral energy distribution (SED) of Mrk 501 during a historically high X-ray state \cite{c554}.
Even while not statistically significant, such a feature could be produced in a number of interesting emission scenarios, such as a two (possibly interacting) zones model, an emission from the magnetosphere of the black hole, or a pile-up in the electron energy distribution due to the stochastic acceleration \cite{c554}. 

\begin{figure}
  \centering
  \includegraphics[width = 0.49\textwidth]{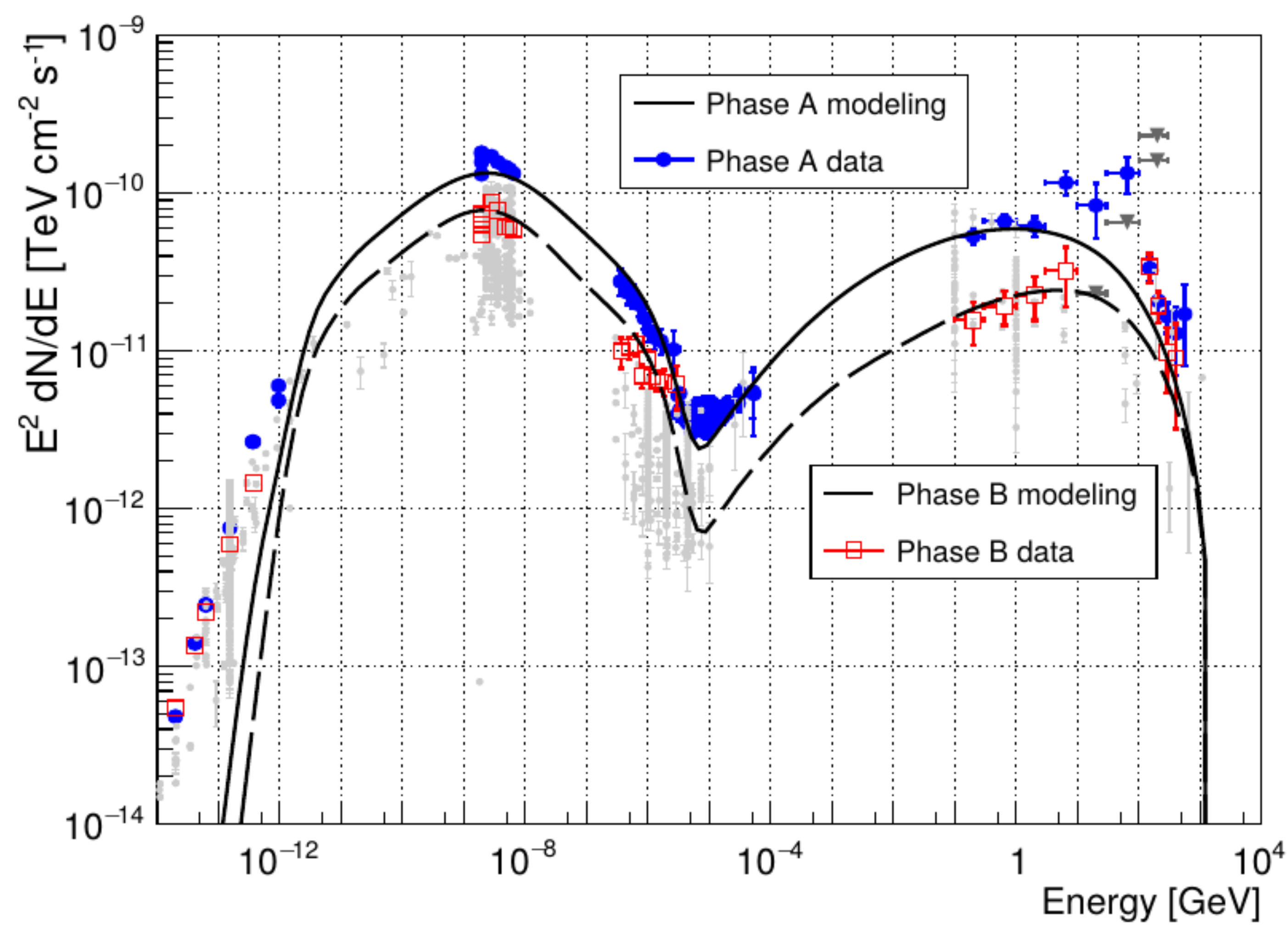}
  \caption{Spectral energy distribution of S5 0716+714 modelled with two zone model \cite{c709}.}
  \label{fig:agn_flares}
\end{figure}

A number of other flares and the theoretical interpretation of the emission has been reported in this conference.
1ES 1959+650 experienced a strong and fast flare with a hard VHE gamma-ray spectrum \cite{c635}.
While the broadband SED can be sufficiently well described by either a leptonic, a lepto-hadronic or a proton-synchrotron model, it requires rather extreme parameters to occur in the source.
A strong flare of Mrk 421 with rich MWL coverage has been observed in 2013 \cite{c624}.
The emission shows a slow raise of the flux over time scale of hours and a faster flare.
A possible explanation is a magnetic reconnection layer moving relativistically within the jet of the source.
The slow raise of the flux would correspond to the combined emission of many plasmoids, while the faster flare would be caused by a single dominant plasmoid (see \cite{c624} and references therein).
A flare with a nightly-scale variability has been seen also from 1ES 1218+304 in 2019 showing a remarkably hard VHE gamma-ray spectrum \cite{c755}.
Two outbursts were observed from the BL Lac object S5 0716+714 in 2015 and 2017 \cite{c709}.
The former outburst showed also a concurrent rotation of EVPA rotation and an appearance of a new knot from the 43 GHz VLBI core.
The possible explanation of the VHE gamma ray emission involves the superluminal knot entering and leaving the recollimation shock in the inner jet (see \cite{c709} and references therein).
The broadband emission can be modelled in a framework of a two-zone scenario (see Fig.~\ref{fig:agn_flares}). 
Also two outbursts from a known VHE gamma ray emitter 3C 279 has been observed in 2017 and 2018 \cite{c668}. 
The level of correlation between the VHE gamma-ray fluxes and the optical emission changes however between the flares making the interpretation of the source more complicated.

\subsection{Radio galaxies}
Radio galaxies, while not as bright as blazars, are normally closer sources, hence we can study the TeV emission without a strong absorption by the EBL.
Moreover, since they are observed at a significant angle to the jet axis, it gives a possibility for imaging of the jet morphology.
The H.E.S.S. Collaboration reported observations of an extended VHE gamma-ray emission from the Cen A radio galaxy.
Nearly 200\,hrs of data combined with careful runwise simulations of the telescope response allowed the detection of a $\sim 2.8$\,kpc projected extension along the direction of the jet axes \cite{c657}. 

The VHE gamma-ray catalog of radio galaxies still contains only a handful of sources. 
The newest addition is 3C 264 \cite{c651}.
The VERITAS observations of this source were motivated by the upcoming interaction of two knots moving with different velocities along the jet.
Despite an elevated flux in optical and X-rays and the detection of a variable VHE gamma-ray emission, no strong VHE gamma-ray flare has been seen, suggesting that the real interaction of the knots might still happen in the near future \cite{c651}. 

\section{VHE gamma-ray follow up of the transients}\label{sec:trans}
In order to obtain possibly the most complete information about the emission of the sources it is important to observe them over broad range of frequencies (multiwavelength, MWL) and even with different messengers (multi messenger, MM).
An important part of such studies are the rapid follow-ups of short electromagnetic alerts (such as Gamma Ray Bursts, GRBs) or observations of sky patches corresponding to the reconstructed directions of selected neutrinos or gravitational wave events.
For the fastest triansients the follow-up must happen automatically, without human intervention. 

\subsection{GRBs}
For a long time the Holy Grail of the transient follow-up studies of the $3^\mathrm{rd}$ generation of the IACTs was to detect the emission from a GRB with such an instrument.
Those events, while extremely bright are also very short, and normally happen at cosmological distances (redshift $\gtrsim 1$), which strongly hinders the observations.
So far of the order of 200 GRB follow-up observations were performed by Cherenkov telescopes, but until recently there had been no reports of the detection of the emission \cite{c634,c761,c782}.

In this conference the detection of the GRB190114C by the MAGIC telescopes was reported. 
Thanks to the relatively low redshift of this event and fast response of the telescopes, despite the not fully favourable observational conditions (large zenith angle and moderate moonlight) strong detection with a statistical significance of over $50\sigma$ was reached in the first few tens of minutes after the burst.
Moreover, the observed flux was very high, reaching 0.1 kCrab level, making it the strongest VHE gamma-ray source observed up to date \cite{c010}.

Furthermore, the H.E.S.S. Collaboration reported the detection of a $5\sigma$ excess in the analysis of data of GRB180720B \cite{c761}.
In this case, the emission has been observed during the afterglow in a time scale of about ten hours.
Interestingly, a $3 \sigma$ hint of an emission also during an afterglow has been seen in the MAGIC data of short GRB160821B, the only source candidate with better observational conditions than GRB190114C \cite{c703}.
Such emission might be explained with a SSC emission from an external shock \cite{c703}.

The GRBs can also be studied with surface arrays. 
In this case, as long as the GRBs are in instantaneous FoV of the instrument the emission can be probed to the very onset of the burst. 
However a higher energy threshold of those instruments comparing to IACTs make the detection more difficult. 
Nevertheless, the brightest GRBs, if they appear in the FoV of HAWC, could be detected \cite{c679}. 

\subsection{Gravitational waves}
In the recent years there is a sudden growth of gravitational wave (GW) astronomy following the successful detection of a number of events \cite{c003}.
The follow up of such events is however very difficult because of the still large uncertainty in the arrival direction.
Therefore clever follow-up algorithms have to be implemented to maximize the exposure in the most probable directions of the alert \cite{c789,c782, c633}.
In this case the large instantaneous FoV of surface array detectors is a strong advantage \cite{c737}, however so far neither Cherenkov telescopes nor surface arrays detected emission coincident with a GW alert.
Interesting limits on the emission have been however obtained by the H.E.S.S. telescopes on the EM170817, the GW alerts associated with a GRB.
The H.E.S.S. observations were performed only 5.5h after the event and also in the next half a year during the ramping up of the radio and X-ray flux and put some constraints on the magnetic field in the merger remnant \cite{c756}. 

\subsection{Neutrino follow-up}
The recent detection of gamma-ray emission from the TXS 0506+056 blazar with the MAGIC telescopes consistent with the arrival direction of the high-energy neutrino IceCube-170922A made this previously anonymous blazar famous and renewed interest in hadronic models of gamma-ray emission \cite{2018Sci...361.1378I}.
In order to characterize the source better, it has become the object of MWL campaigns and VHE gamma-ray monitoring.
The source has been re-detected by VERITAS, albeit at lower flux then the one detected by MAGIC, over longer time scale following the original flare \cite{c632}.
Interestingly, the monitoring by MAGIC performed in 2018/2019 was rewarded with a detection of another flare from this object and a hint of an emission in a low state \cite{c646}.
The VHE gamma-ray emission during the flare is very similar to the 2017 event, and can be again explained by mostly leptonic processes, which is also consistent with lack of neutrino events observed during the newer flare \cite{c646}. 

\begin{figure}
  \centering
  \includegraphics[width = 0.53\textwidth]{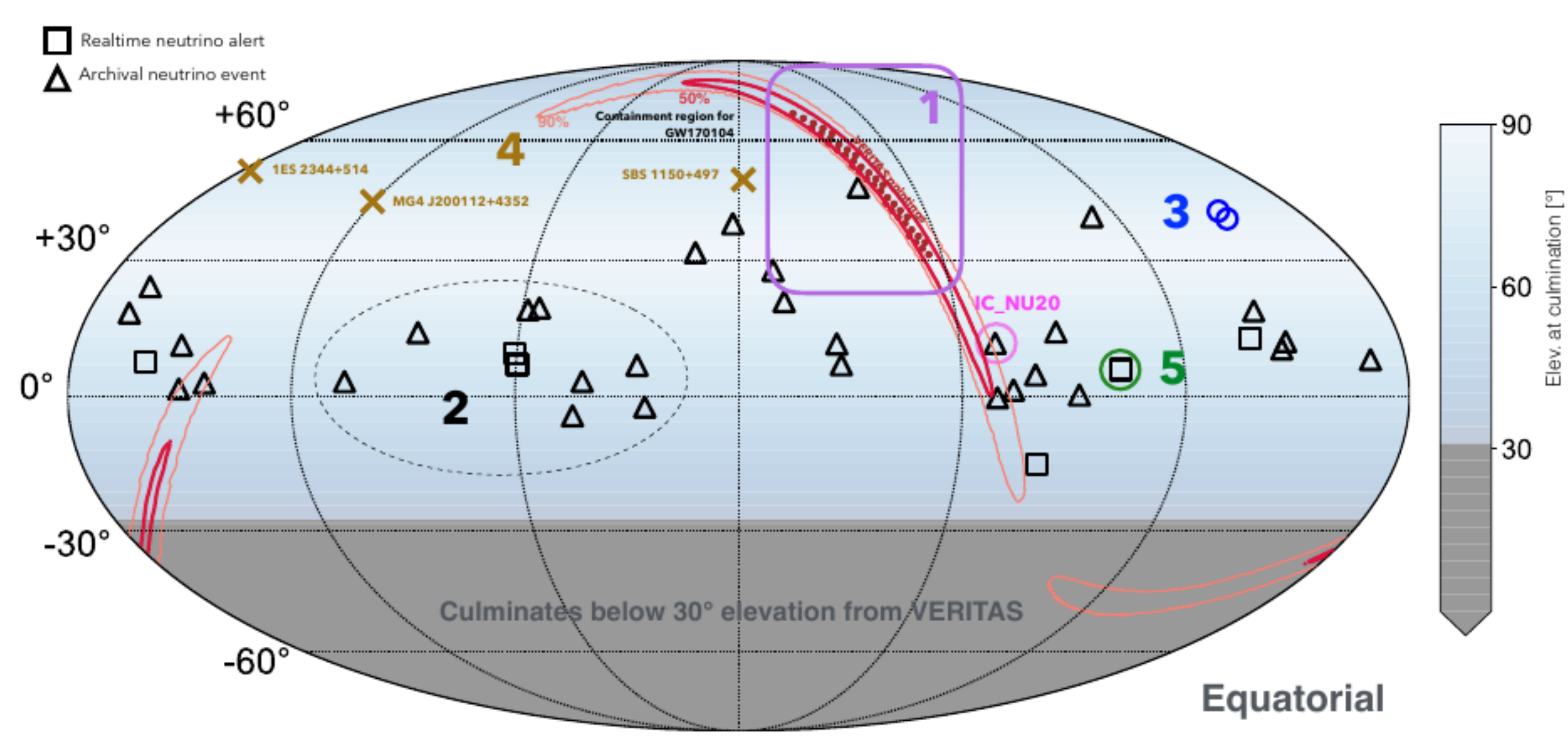}
  \includegraphics[width = 0.46\textwidth]{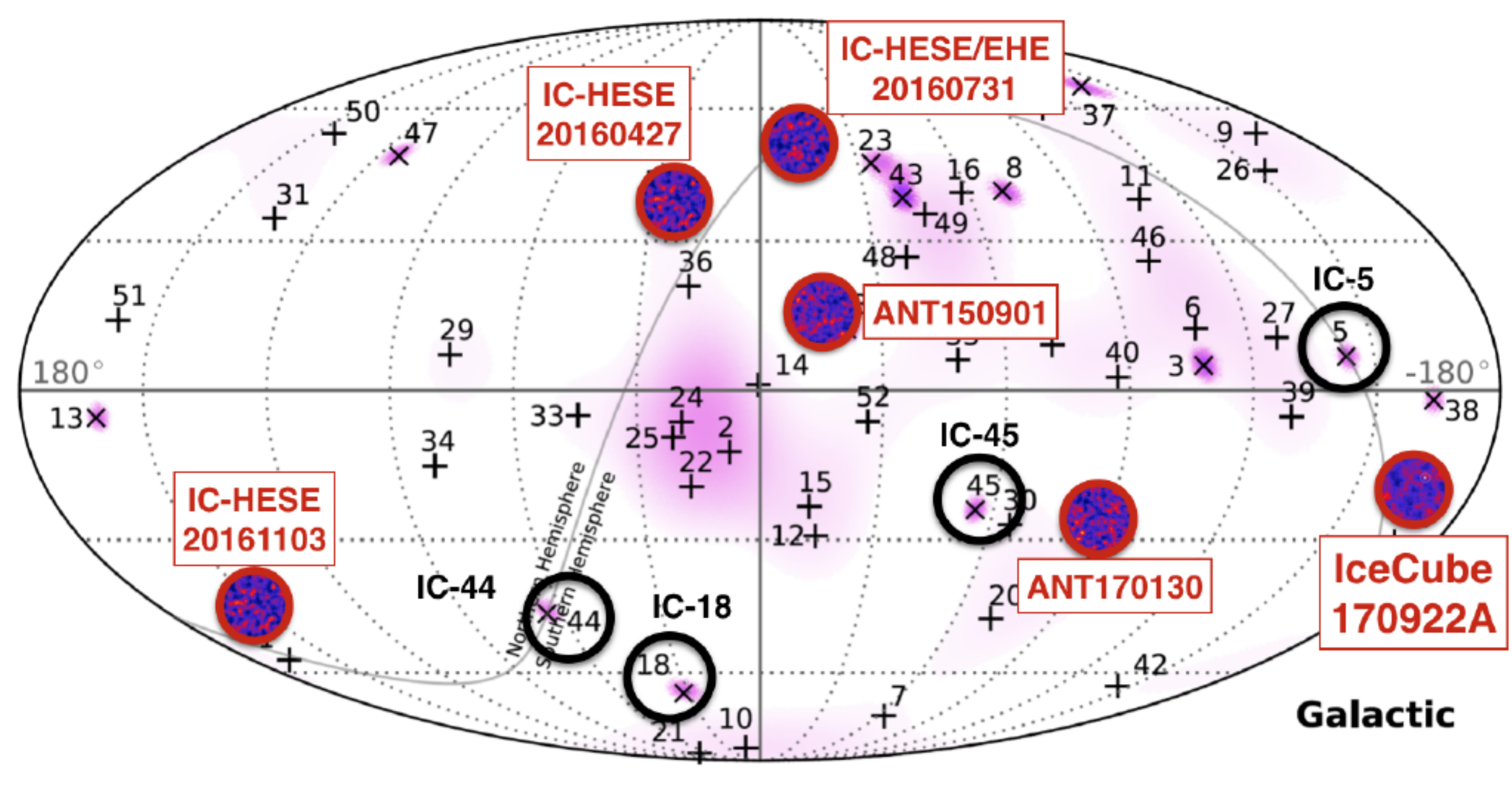}
  \includegraphics[width = 0.46\textwidth]{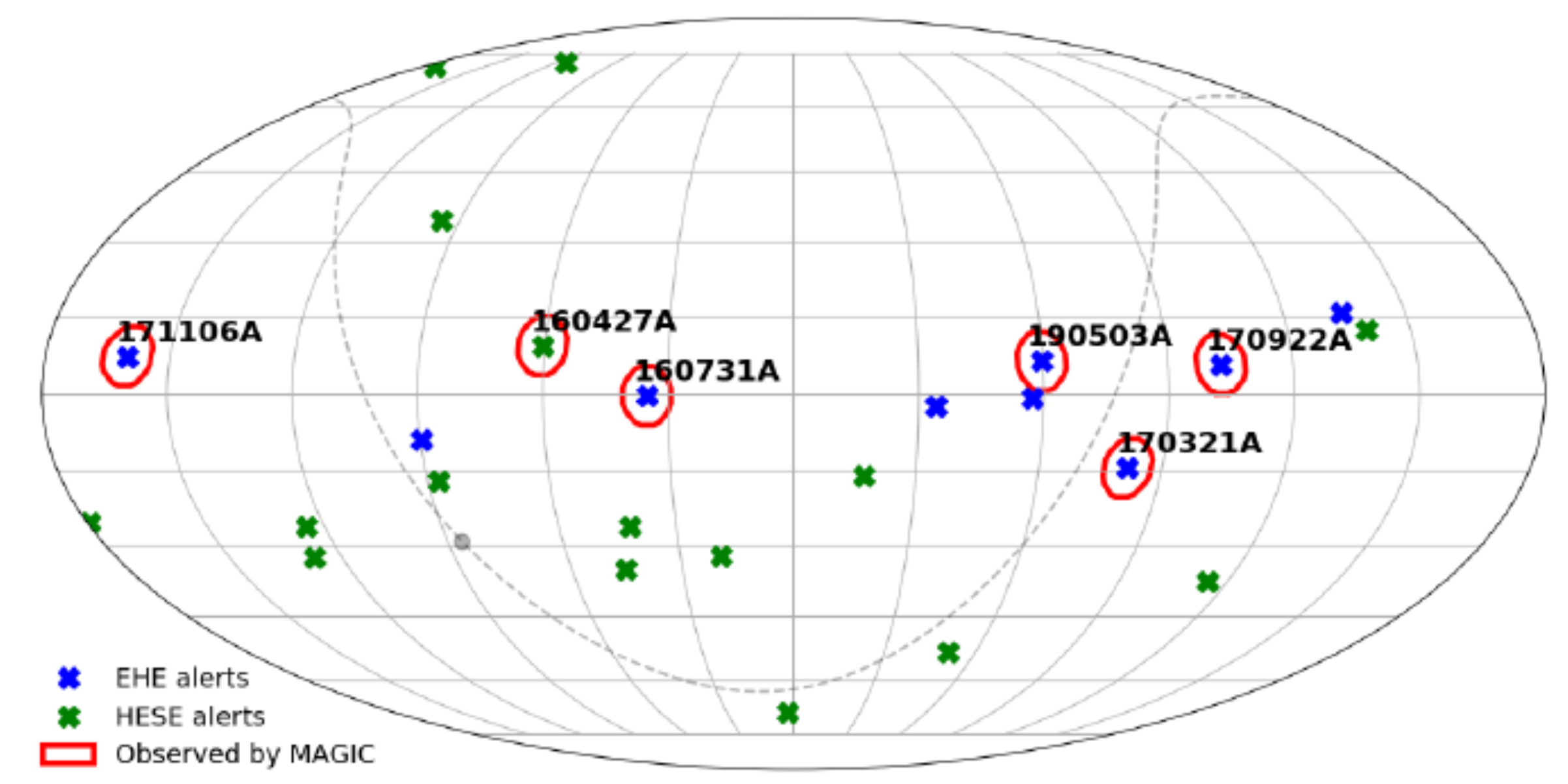}
  \caption{Sky map showing the followed multimessenger triggers by VERITAS (left panel, equatorial coordinates, \cite{c782}), H.E.S.S. (right panel, galactic coordinates, \cite{c787}) and MAGIC (bottom panel, equatorial coordinates, \cite{c671}.}
  \label{fig:mm_followup}
\end{figure}

The search for other ``neutrino blazars'' continues and new promising neutrino events are being followed \cite{c782, c633, c787} (see Fig.~\ref{fig:mm_followup}), however so far no other VHE gamma-ray flare has been detected associated to another neutrino event. 

\subsection{FRB follow-up}\label{sec:frb}
Fast radio bursts are short (ms-duration), bright flashes of radio emission of extragalactic origin.
So far their origin is unknown.
Most of them are one time only events, however two repeating sources have been seen: FRB 121102 and FRB 190814.J0422+73.
To shed light on the origin of FRB gamma-ray observations of FRB repeaters are being performed \cite{c633, c698}. 
Due to shortness of the phenomena, observations have to be performed strictly simultaneously with a radio telescope, and the coincidence of a possible signal with FRB is done offline.
Interestingly, Cherenkov telescopes, due to their huge mirror areas can also be used as optical telescopes for looking for optical counterparts of FRBs \cite{c007}.
While during the observations times with the Cherenkov telescopes a number of FRB have been detected by radio telescopes, no significant emission either as gamma-ray excess, nor raise of optical flux has been observed \cite{c698, c633}.

\section{Exotic sources}\label{sec:exotic}
In this section various more ``exotic'' sources are discussed, from which the TeV emission is being searched with Cherenkov telescopes or surface arrays.

\subsection{The Sun}
GeV emission has been detected by \textit{Fermi}-LAT from the Sun during its solar minimum \cite{2011ApJ...734..116A}.
Observations with HAWC so far have not resulted in the detection of a TeV extension of such an emission.
The data taken in 2018, during falling activity of the Sun are above the extrapolation of the \textit{Fermi}-LAT measurement  \cite{c369}.

\subsection{UHE gamma rays}
As discussed in Section~\ref{sec:100TeV} the highest energies of detected gamma rays touch hundreds of TeV.
Search of much higher, of the order of EeV, energies is being performed with instruments normally used for studies of CRs (CR surface arrays and hybrid detectors) using differences in the characteristics of CR and gamma ray induced showers \cite{c326, c398}.
Nevertheless so far no such emission has been detected and the limits are still at least a factor of a few away from the model predictions of the diffuse EeV gamma-ray flux \cite{c398}.  

Gamma rays of such energies can be affected by the so-called preshower effect, i.e. interaction between the EeV gamma ray and the Earth's magnetic field.
This can both influence the existing searches for EeV gamma rays, as well as provide an alternative method for probing this energy range \cite{c688}.

Also radio detectors can be used to study the VHE gamma-ray emission of sources down to energies as low as 100\,TeV if sufficient number of mini-arrays is used \cite{c655}.
While the method is still in proof of principle state, and there are still many unknowns in this technique, including e.g. the effect of the background from hadronic showers \cite{c655}, it is reassuring that the estimated range of sensitivity starts to overlap with the highest energies seen from Galactic sources (see Section~\ref{sec:100TeV}). 

\subsection{Probing fundamental physics and cosmology with gamma rays}
Detection of new, exotic classes of gamma ray sources can also influence fundamental physics.
Moreover, the emission from known sources can also be used to probe cosmology.
An example of the former is the search for evaporating primordial black holes (PBH).
The PBH with masses slightly exceeding $5 \times 10^{14}$\,g would evaporate now, producing gamma rays in their final moments \cite{c719, c804}.
Since the PBH emission can come from an arbitrary direction in the sky, surface arrays like HAWC \cite{c516} can use their whole FoV, and the Cherenkov telescopes  \cite{c719, c804} can stack thousands of hours of observations of different FoVs.
While no detection has been claimed yet, competitive upper limits on the PBH evaporation density can be put.

\begin{figure}
  \centering
  \includegraphics[width = 0.6\textwidth]{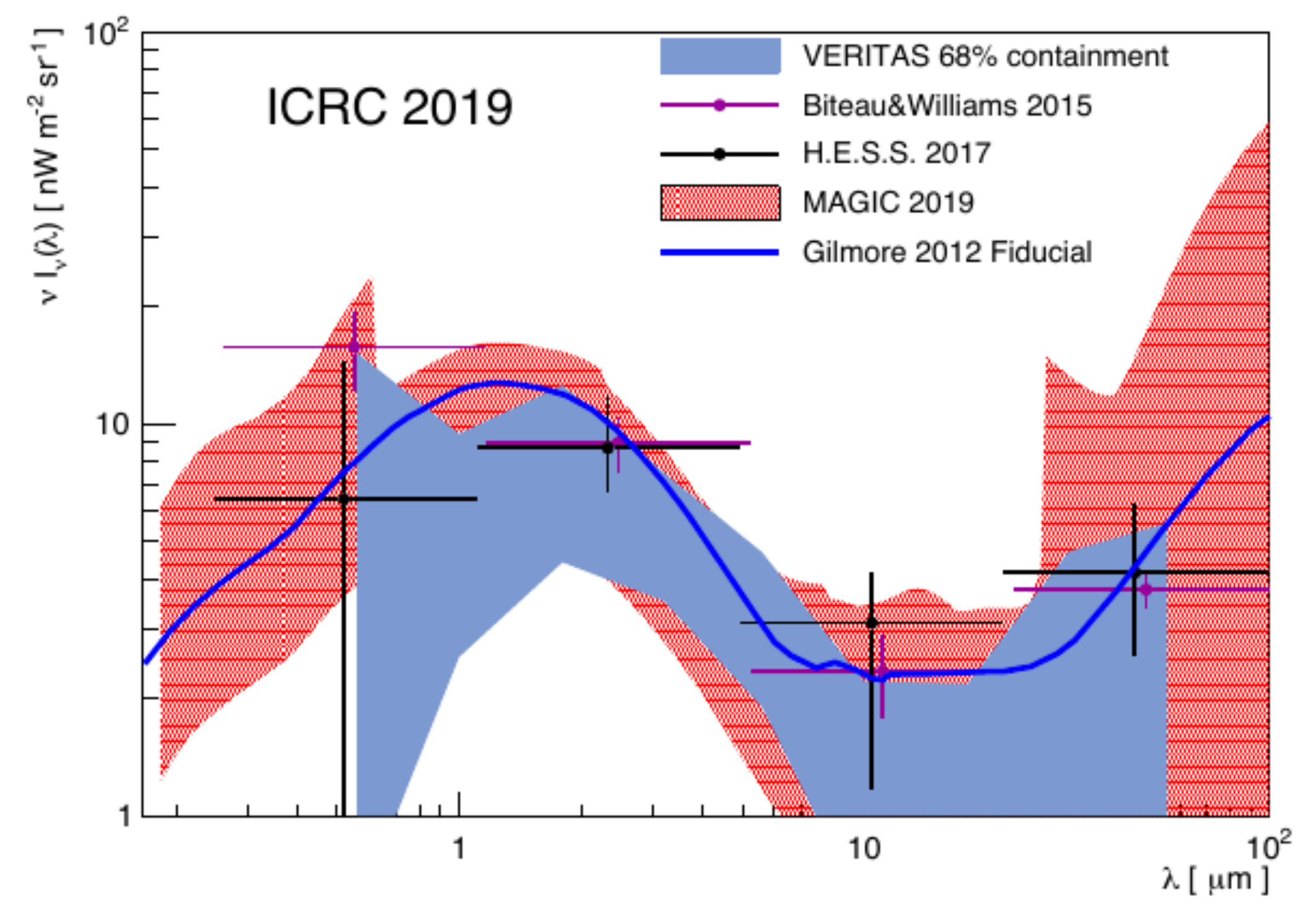}
  \caption{Constraints on EBL density obtained with various IACTs \cite{c770}.
  Note that the treatment of systematic uncertainties varies between different experiments. }
  \label{fig:ebl}
\end{figure}

The propagation of gamma rays from extragalactic sources to the observer can also be used to probe the content of the Universe.
The observations of the effect on the spectra of blazars from the absorption of gamma rays by the EBL have become over the last decade a well grounded method for the estimation of the EBL. 
Not only new data provide more precise statistical uncertainties, but also inclusion of more refined systematic effect studies makes the results more robust (see e.g. \cite{2017A&A...606A..59H, 2019MNRAS.486.4233A} and references therein). 
In the conference the updated results from the VERITAS Collaboration have been presented \cite{c770}. 
The EBL constraints obtained with different experiments agree with each other (see Fig.~\ref{fig:ebl}) and with other methods.
They also agree with the recent EBL models. 
However, the EBL is still poorly constrained at the lowest and highest wavelengths, which can be probed by detection of emission from the $z\gtrsim 1$ blazars and by observations of the tens of TeV energies of the closest blazars, respectively. 

The absorption of gamma rays from extragalactic sources by the EBL can be used also to probe the intergalactic magnetic fields (IGMF), see references within \cite{c763}.
$e^+e^-$ pairs produced in this process can then inverse Compton scatter the cosmic microwave background photons to produce a secondary gamma-ray emission.
Depending on the strength of IGMF in which the propagation of the electrons is occuring, such secondary emission would be observable either from the same direction (resulting only in modified SED), from a somewhat extended source, or nearly isotropic.
In the recent years the observations both with IACTs and with \textit{Fermi}-LAT considerably limited the phase space of the strength and correlation length of IGMFs using this effect.
No detection of AGN halos (despite some early claims) and the shrinking phase space however raised a question if another energy loss process, such as plasma instabilities \cite{c763} can be competing with the production of secondary gamma rays, preventing the formation of observable AGN halos.

The propagation of gamma rays can also be used to test Lorentz Invariance Violation (LIV), i.e. dependence of the speed of light on the energy of photons.
The most direct method involves searching for energy-dependent delays in fast varying sources observed at large distances up to the highest possible energies.
This makes sources like AGN flares, GRBs and pulsars interesting targets for LIV searches (see e.g. \cite{2013PhRvD..87l2001V}).
However, no update on employing such a method has been presented in this conference. 

For the surface arrays the performance of this method is however hindered somewhat by the low sensitivity for short time scales.
Nevertheless, in the case of a superluminal LIV another effect can be exploited.
The highest energy photons then can decay, thus observations of $>100$\,TeV gamma rays can put competitive limits on LIV \cite{c738}.
LIV effects can also affect the absorption of the gamma rays in the EBL.
Exploiting this effect allows putting stringent constraints on the LIV in the subluminal case using a large collection of over a hundred measured spectra from 38 known TeV sources \cite{c658}. 

\section{Progress in experimental methods}\label{sec:methods}
Many contributions of the conference were devoted to development of analysis methods, in particular for the upcoming experiments (mainly CTA, but also e.g. LHAASO).
In this section I highlight a few of those that have been already successfully implemented in the existing experiments. 

\subsection{Open Cherenkov Astronomy}
With the onset of the CTA Observatory not only higher quality data are expected but also a more open access both to the data and to the analysis tools of Cherenkov telescopes \cite{c717}.
This openness also starts to sink into the current generation of Cherenkov telescopes.
The H.E.S.S. Collaboration has released a data set of about 2500\,hr of the Galactic plane survey, and another release of a similar amount of extragalactic data is planned \cite{c656}.
The MAGIC Collaboration at the moment is providing mostly only the machine readable high-level products of the analysis, however in some cases event lists were provided and there is work ongoing on providing a legacy dataset in CTA-like DL3 format \cite{c666}.

An interesting idea of exploiting Citizen Science has been employed by the VERITAS Collaboration.
Using a popular portal a large set of Cherenkov images of events has been posted with the request for classification to a large sample of volunteer classifiers.
Originally the idea involved separation of gamma-rays from background, however it proved to be much more successful in the identification of muon images \cite{c678}.

\subsection{Observations at very high zenith angles}
In order to perform observations at the highest energies a huge collection area is required due to very low fluxes of sources at such energies.
This can be achieved by making large, sparsely distributed arrays of telescopes (as e.g. SST in CTA, \cite{c741}).
However a large collection area can be also achieved with the current generation of telescopes exploiting observations of very large ($\gtrsim 70^\circ$) zenith angles.
In such a case the shower maximum happens far away from the telescope and thus the Cherenkov light is distributed over a very large area.
This results in a higher collection area at the highest energies at the price of an increased energy threshold and worse low energy performance of the instrument. 
The technique has a number of intrinsic difficulties, such as it requires careful monitoring of atmospheric transparency, produces vertical acceptance gradient along the camera, results in more compact Cherenkov images affecting gamma/hadron separation.
Nevertheless, it was successfully applied to the MAGIC telescopes in the search of photons with energies of the order of 100\,TeV \cite{c828}.
It also allowed the study of variability of the Crab Nebula at energies above 10 TeV on time scales of months \cite{c812}. 

\subsection{Stellar Intensity Interferometry with IACTs}
Comparing to the classical optical telescopes, Cherenkov telescopes have much larger light collection area, but much worse optical point spread function.
The first characteristic makes them a useful instrument to study very short optical flares, such as possible counterparts of FRBs (see Section~\ref{sec:frb} and \cite{c007}).

An even more interesting application is the Stellar Intensity Interferometry (SII).
Its principle involves simultaneous, continuous observations of an extended optical source with at least two telescopes and calculating the correlation of the signal obtained in both.
The level of correlation depends on the baseline length and on the angular size of the source. 
Application of this method made possible for VERITAS telescopes to measure diameters of two stars \cite{c740}. 
The full power of SII with Cherenkov telescopes will be released only once CTA is constructed. 
Larger number of telescopes, spread over a large area, will provide simultaneous sensitivity to various angular scales making possible to get an actual image of the star \cite{c007}. 

\subsection{Surface arrays analysis methods}
Search for the highest energies observable with HAWC motivated the usage of artificial neural networks in the energy estimation of the events resulting in a significant improvement over the original energy estimation technique used in HAWC \cite{c723}.
In addition, an improvement in the HAWC performance is expected soon, when the extra information of the outrigger array is incorporated in the analysis \cite{c736}

\section{Summary}\label{sec:sum}
A number of very important results has been reported in this conference.
The GRB physics can be finally probed above a few hundreds of GeV.
Several sources have been reported to emit at or beyond 100\,TeV, and are candidates for Cosmic Ray PeVatrons.
Two small populations of VHE gamma-ray sources, from which before only individual sources were known, start to appear: pulsars and EHBLs.
Jets of both a radio galaxy and a microquasar have been resolved in TeV gamma rays.
It is expected that with the onset of CTA many more interesting results will come in the nearest future. 

\section*{Acknowledgements}
This work is supported by the grant through the Polish Narodowe Centrum Nauki No. \\2015/19/D/ST9/00616.
JS would like to thank R. L\'opez-Coto and M. Cerruti for providing comments to the manuscript.


\begin{thebibliography}{999}
\bibitem{c656} M. De Naurois, PoS(ICRC2019)656 
\bibitem{2016APh....72...61A} Aleksi{\'c}, J., Ansoldi, S., Antonelli, L.~A., et al.\ 2016, Astroparticle Physics, 72, 61 
\bibitem{c632} W. Benbow, PoS(ICRC2019)632  
\bibitem{c773} G. Richards, PoS(ICRC2019)773  
\bibitem{c665} D. Dorner, A. Arbet-Engels, D. Baack,  et al., PoS(ICRC2019)665 
\bibitem{c741} D. Mazin, PoS(ICRC2019)741 
\bibitem{c653} J. Cortina for the CTA LST project,  PoS(ICRC2019)653
\bibitem{c015} K. Tollefson, PoS(ICRC2019)015, presentation 
\bibitem{c736} V. Marandon, A. Jardin-Blicq and H. Schoorlemmer,  PoS(ICRC2019)736 
\bibitem{c778} T. Sako, PoS(ICRC2019)778, presentation 
\bibitem{c785} H. Schoorlemmer, PoS(ICRC2019)785 
\bibitem{c786} F. Sch\"ussler, PoS(ICRC2019)786, 
\bibitem{c693} H. He, PoS(ICRC2019)693 
\bibitem{c834} J. Zorn, PoS(ICRC2019)834 
\bibitem{c802} D. Strom, PoS(ICRC2019)802 
\bibitem{c828} I. Vovk, M. Will, R. Mirzoyan, J. Besenrieder, M. Peresano, P. Temnikov, J. van Scherpenberg and D. Zaric,  PoS(ICRC2019)828 
\bibitem{c714} D. Kieda, for the VERITAS Collaboration, S. LeBohec and R. Cardon, PoS(ICRC2019)714 
\bibitem{c393} T. Capistr\'an Rojas, I.D. Torres Aguilar, E. Moreno Barbosa and on behalf of the HAWC Collaboration, PoS(ICRC2017)393 
\bibitem{c697} R. Parsons, P. Bordas, S. Klepser and H.E.S.S. Collaboration, PoS(ICRC2017)697 
\bibitem{c706} A. Jardin-Blicq, V. Marandon and F. Brun,  PoS(ICRC2019)706 
\bibitem{2016Natur.531..476H} HESS Collaboration, Abramowski, A., Aharonian, F., et al.\ 2016, Nature, 531, 476 
\bibitem{c680} C. Fruck, I. Vovk, M. Strzys, Y. Iwamura and S. Ventura, PoS(ICRC2019)680 
\bibitem{c816} S. Ventura, D. Grasso and A. Marinelli, PoS(ICRC2019)816 
\bibitem{tevcat} \url{http://tevcat.uchicago.edu/}
\bibitem{c801} C. Steppa, PoS(ICRC2019)801 (presentation). 
\bibitem{c799} M. Spir-Jacob, A. Djannati-Atai, L. Mohrmann, G. Giavitto, B. Khelifi, B. Rudak, C. Venter and R. Zanin, PoS(ICRC2019)799 
\bibitem{c728} M. L\'opez Moya, PoS(ICRC2019)728 
\bibitem{2017Sci...358..911A} Abeysekara, A.~U., Albert, A., Alfaro, R., et al.\ 2017, Science, 358, 911 
\bibitem{c832} H. Zhou, PoS(ICRC2019)832
\bibitem{c020} G. Giacinti, PoS(ICRC2019)020 (presentation) 
\bibitem{c670} K. Fang, X.J. Bi and P.F. Yin, PoS(ICRC2019)670 
\bibitem{c685} G. Giacinti and R. Lopez-Coto, PoS(ICRC2019)685 
\bibitem{c797} A. Smith, PoS(ICRC2019)797 
\bibitem{c640} C. Brisbois and H. Zhou, PoS(ICRC2019)640 
\bibitem{c809} M. Tsirou, Z. Meliani and Y. Gallant, PoS(ICRC2019)809 
\bibitem{c715} N. Komin and H.E.S.S. Collaboration, PoS(ICRC2019)715 (presentation) 
\bibitem{c639} C. Brisbois and V. Joshi, PoS(ICRC2019)639 
\bibitem{c827} D. Zaric, R. Mirzoyan, I. Vovk, P. Temnikov, M. Peresano, N. Godinovi\'c, J. van Scherpenberg and J. Besenrieder, PoS(ICRC2019)827 
\bibitem{c674} H. Fleischhack, PoS(ICRC2019)674
\bibitem{c564} D. Green, T. Nagayoshi and F. dePalma, PoS(ICRC2019)564 
\bibitem{2004ApJ...614..897A} Aharonian, F., Akhperjanian, A., Beilicke, M., et al.\ 2004, ApJ, 614, 897 
\bibitem{2006A&A...457..899A} Aharonian, F., Akhperjanian, A.~G., Bazer-Bachi, A.~R., et al.\ 2006, A\&A, 457, 899 
\bibitem{c759} M. Peresano, R. Mirzoyan, I. Vovk, P. Temnikov, D. Zaric, N. Godinovi\'c, J. van Scherpenberg, J. Besenrieder for the MAGIC Collaboration, PoS(ICRC2019)759 
\bibitem{c734} K. Malone, PoS(ICRC2019)734 
\bibitem{c712} K. Kawata for  Tibet AS gamma Collaboration,  PoS(ICRC2019)712 
\bibitem{c723} J. Linnemann, P. Harding, J. Lundeen, S. Marinelli and H. Mart\'inez-Huerta, PoS(ICRC2019)723 
\bibitem{c732} G. Maier, O. Blanch, D. Hadasch, N. Komin, M. Lundy, A. L\'opez-Oramas, D. Malyshev, J. Moepi, S. Ohm, G. P\"uhlhofer, R. Prado, S. Schlenstedt, D.F. Torres and B. Zitze, PoS(ICRC2019)732 
\bibitem{c713} D. Kieda for VERITAS Collaboration, PoS(ICRC2019)713 
\bibitem{c824} T. Williamson, PoS(ICRC2019)824 
\bibitem{c813} B. van Soelen, PoS(ICRC2019)813, presentation 
\bibitem{c1178} I. Sushch,  PoS(ICRC2019)1178, presentation 
\bibitem{c772} C.D. Rho, C.D. Rho, H. Zhou and S. BenZvi, PoS(ICRC2019)772 
\bibitem{c699} B. Hona, H. Fleischhack and P. Huentemeyer, PoS(ICRC2019)699 
\bibitem{c716} N. Komin, M. Haupt and H.E.S.S. Collaboration, PoS(ICRC2019)716
\bibitem{c781} F. Salesa Greus and S. Casanova, PoS(ICRC2019)781 
\bibitem{c682} J.A. Garcia-Gonzalez, M.M. Gonz\'alez and N. Fraija, PoS(ICRC2019)682 
\bibitem{c796} V. Sliusar, A. Arbet-Engels, D. Baack, et al., PoS(ICRC2019)796 
\bibitem{2019MNRAS.487.3990B} Bhatta, G.\ 2019, MNRAS, 487, 3990 
\bibitem{c629} J. Becerra Gonzalez, J. Sitarek, C. Nigro, E. Lindfors, F. Tavecchio, V.F. Ramazani for the  MAGIC and \textit{Fermi}-LAT collaborations, PoS(ICRC2019)629 
\bibitem{c768} E. Prandini, C.H.E. Arcaro, K. Asano, et al., PoS(ICRC2019)768 
\bibitem{c676} L. Foffano, E. Prandini, A. Franceschini and S. Paiano, PoS(ICRC2019)676 
\bibitem{c620} A. Arbet-Engels, M. Manganaro, D. Dorner, et al., PoS(ICRC2019)620 
\bibitem{c822} T. Weisgarber, PoS(ICRC2019)822 
\bibitem{c689} O. Gueta, PoS(ICRC2019)689 
\bibitem{c554} J. Becerra Gonzalez, D. Paneque, C. Wendel, F. Tavecchio, K. Noda, K. Ishio, J. Sitarek for the MAGIC and \textit{Fermi}-LAT  collaborations, PoS(ICRC2019)554 
\bibitem{c635} W. Bhattacharyya, M. Takahashi, M. Hayashida and on behalf of the \textit{Fermi}-LAT and MAGIC collaborations,   PoS(ICRC2019)635 
\bibitem{c624} A. Babic, T. Hassan, D. Paneque, M. Balokovic, J. Finke and M. Petropoulou, PoS(ICRC2019)624 
\bibitem{c755} S. O Brien, PoS(ICRC2019)755 
\bibitem{c709} Y. Kajiwara, M. Manganaro, E. Lindfors, B. Rani, S.G. Jorstad, V.M. Larionov and A. Marscher, PoS(ICRC2019)709 
\bibitem{c668} G. Emery, M. Cerruti, A. Dmytriiev, F. Jankowsky, H. Prokoph, C. Romoli and M. Zacharias, PoS(ICRC2019)668 
\bibitem{c657} M. De Naurois, PoS(ICRC2019)657, presentation 
\bibitem{c651} J. Christiansen, PoS(ICRC2019)651, presentation 
\bibitem{c634} A. Berti, L.A. Antonelli, Z. Bosnjak, et al., PoS(ICRC2019)634  
\bibitem{c761} Q. Piel, C.H.E. Arcaro, H. Ashkar, et al., PoS(ICRC2019)761 
\bibitem{c782} M. Santander, PoS(ICRC2019)782 
\bibitem{c010} R. Mirzoyan, PoS(ICRC2019)010, presentation 
\bibitem{c703} S. Inoue, L. Nava, K. Noda, et al., PoS(ICRC2019)703 
\bibitem{c679} N. Fraija and M.M. Gonzalez, PoS(ICRC2019)679 
\bibitem{c003} E. Bissaldi, PoS(ICRC2019)003 
\bibitem{c789} M. Seglar-Arroyo, H. Ashkar, S. Bonnefoy, et al., PoS(ICRC2019)789 
\bibitem{c633} A. Berti, E. Bernardini, W. Bhattacharyya, et al., PoS(ICRC2019)633 
\bibitem{c737} I. Martinez, PoS(ICRC2019)737 
\bibitem{c756} S. Ohm, PoS(ICRC2019)756 
\bibitem{2018Sci...361.1378I} IceCube Collaboration, Aartsen, M.~G., Ackermann, M., et al.\ 2018, Science, 361, eaat1378 
\bibitem{c646} M. Cerruti, E. Bernardini, W. Bhattacharyya, et al., PoS(ICRC2019)646, presentation 
\bibitem{c787} F. Sch\"ussler, H. Ashkar, M. Backes, et al., PoS(ICRC2019)787 
\bibitem{c671} A. Fattorini, PoS(ICRC2019)671 
\bibitem{c698} J. Holder for the VERITAS Collaboration and R.S. Lynch, PoS(ICRC2019)698 
\bibitem{c007} M. Daniel, PoS(ICRC2019)007 
\bibitem{2011ApJ...734..116A} Abdo, A.~A., Ackermann, M., Ajello, M., et al.\ 2011, ApJ, 734, 116 
\bibitem{c369} M. Nisa, PoS(ICRC2019)369 
\bibitem{c326} M. Kuznetsov, O. Kalashev and G. Rubtsov, PoS(ICRC2019)326 
\bibitem{c398} J. Rautenberg, PoS(ICRC2019)398 
\bibitem{c688} D. Gora, K. Almeida Cheminant, N. Dhital, P. Homola, O. Sushchov and J. Zamora-Saa, PoS(ICRC2019)688 
\bibitem{c655} R. Dallier, L. Bondonneau, D. Charrier, et al., PoS(ICRC2019)655 
\bibitem{c719} S. Kumar, PoS(ICRC2019)719 
\bibitem{c804} T. Tavernier, J.F. Glicenstein and F. Brun, PoS(ICRC2019)804 
\bibitem{c516} K. Engel, A. Peisker, P. Harding, J. Wood, I. Martinez-Castellanos, A. Albert and K. Tollefson, PoS(ICRC2019)516, 
\bibitem{2017A&A...606A..59H} H.~E.~S.~S.~Collaboration, Abdalla, H., Abramowski, A., et al.\ 2017, A\&A, 606, A59 
\bibitem{2019MNRAS.486.4233A} Acciari, V.~A., Ansoldi, S., Antonelli, L.~A., et al.\ 2019, MNRAS, 486, 4233 
\bibitem{c770} E. Pueschel, PoS(ICRC2019)770 
\bibitem{c763} M. Pohl, PoS(ICRC2019)763 
\bibitem{2013PhRvD..87l2001V} Vasileiou, V., Jacholkowska, A., Piron, F., et al.\ 2013, Phys. Rev. D, 87, 122001 
\bibitem{c738} H. Mart\'inez-Huerta, S. Marinelli, J. Linnemann and J. Lundeen, PoS(ICRC2019)738 
\bibitem{c658} H. Mart\'inez-Huerta, V. de Souza and R.G. Lang, PoS(ICRC2019)658 
\bibitem{c717} K. Kosack and M. Peresano, PoS(ICRC2019)717 
\bibitem{c666} M. Doro, C. Nigro, E. Prandini, A. Tramacere, M. Delfino, J. Delgado, E. do Souto, L. Jouvin and J. Rico, PoS(ICRC2019)666
\bibitem{c678} M. Laraia, D. Wright, H. Dickinson, A. Simenstad, K. Flanagan, S. Serjeant, L. Fortson for VERITAS Collaboration, PoS(ICRC2019)678 
\bibitem{c812} J. van Scherpenberg, R. Mirzoyan, I. Vovk, M. Peresano, D. Zaric, P. Temnikov, N. Godinovi\'c and J. Besenrieder, PoS(ICRC2019)812 
\bibitem{c740} N. Matthews, PoS(ICRC2019)740 
\end{thebibliography}
\end{document}